\documentclass[useAMS,usenatbib,onecolumn]{mn2e}
\usepackage{epsfig}
\usepackage{amsmath, amssymb,bm}

\title[Tidal Truncation of Circumplanetary Discs]{Tidal Truncation of
  Circumplanetary Discs}

\author[R. G. Martin \& S. H. Lubow]{ Rebecca G. Martin and
 Stephen H. Lubow\thanks{E-mail: rmartin@stsci.edu; lubow@stsci.edu}\\
Space Telescope Science Institute, 3700 San Martin
  Drive, Baltimore, MD 21218, USA}

\begin{document}

\pagerange{\pageref{firstpage}--\pageref{lastpage}} 
\pubyear{2010}
\maketitle

\label{firstpage}

\begin{abstract}
We analyse some properties of circumplanetary discs.  Flow through
such discs may provide most of the mass to gas giant planets, and such
discs are likely sites for the formation of regular satellites.  We
model these discs as accretion discs subject to the tidal forces of
the central star.  The tidal torques from the star remove the disc angular momentum
near the disc outer edge and permit the accreting disc gas to lose
angular momentum at the rate appropriate for steady accretion.
Circumplanetary discs are truncated near the radius where periodic
ballistic orbits cross, where tidal forces on the disc are
strong. This radius occurs at approximately $0.4 \, r_{\rm H}$ for the
planet Hill radius $r_{\rm H}$. During the T Tauri stage of disc
accretion, the disc is fairly thick with aspect ratio $H/r \ga 0.2$
and the disc edge tapering occurs over a radial scale $\sim H \sim 0.1
r_{\rm H}$.  The disc fluid equations can be rescaled in the Hill
approximation to a form similar to the flow equations for a disc in a
binary star system with a mass ratio of unity.  For a circular or
slightly eccentric orbit planet, no significant resonances lie within
the main body of the disc.  Tidally driven waves involving resonances nonetheless play an important
role in truncating the disc, especially when it is fairly thick.  We
model the disc structure using one dimensional time-dependent and
steady-state models and also two dimensional SPH simulations.  The
circumplanetary disc structure depends on the variation of the disc
turbulent viscosity with radius and is insensitive to the angular
distribution of the accreting gas.  Dead zones may occur within the
circumplanetary disc and result in density structures.  If the disc is
turbulent throughout, the predicted disc structure near the location
of the regular Jovian and Saturnian satellites is smooth with no
obvious feature that would favor formation at their current locations.
It may be possible that substructure, such as due to variations in the
disc turbulence, could lead to the trapping of migrating satellites.
\end{abstract}


\section{Introduction}

In the core accretion model of planet formation, at the earliest
stages of giant planet formation, the planet is fully embedded in the
gaseous disc that orbits the central star \citep*{mizuno80,pollack96,
  hubickyj07, papaloizou05}. At later stages,
during run-away-gas accretion, tidal forces due to the planet open a
gap in the disc, typically when the planet reaches a mass of order
Neptune's mass \citep*{lin86,bate03,dangelo02}.  The gap opening does
not necessarily imply that the gas flow on to the planet has ceased
\citep{artymowicz96}. The gap structure is determined by the
properties of dynamical flow within it. The relatively low density of
gas in the gap results from the relatively high flow radial velocities
towards the star as a consequence of the disturbance caused by the
planet.  Outside the gap region, the  gas radial drift occurs on the
much slower viscous timescale, resulting in much higher densities than
occur in the gap.  Some studies suggest that most of the inflowing gas
just outside the gap is accreted by a planet whose mass is comparable
to or less than Jupiter's for typical disc parameters
\citep*{bryden99,kley99,lubow99,lubow06}.

During this gap phase, the planetary radius is typically much smaller
than its Hill (tidal) radius. 
Circumplanetary discs may form as gas flowing through the gap on to a
planet carries some angular momentum about the planet
\citep{lubow99,bate03,dangelo02}.  As a result, the inflowing gas has
too much angular momentum to directly strike the planet and instead
forms a disc about it.  Most of the mass of Jupiter may have been
acquired in the gap stage through gas flow that involves a
circumplanetary disc.

There are at least two motivations for studying circumplanetary discs.
The observational detection of a disc orbiting an extra-solar gas
giant planet would provide important evidence about the process of
planet formation. The determination of their expected properties
provides constraints for their detection.  A second motivation is to
better understand satellite formation, since circumplanetary discs are
likely sites of satellite formation.  Satellites that orbit
Jupiter and Saturn are classified into two groups. The regular
satellites have low eccentricity and low orbital inclination to the
equatorial plane of their planet. They are thought to have formed in
the circumplanetary disc \citep{lunine82, canup02, mosqueira03}.  The
second type, the irregular satellites, can have a high eccentricity
and high orbital inclinations. They can orbit either progradely or
retrogradely with respect to the planet spin \citep{grav03}. The circumplanetary disc 
may capture these satellites though gas drag \citep*{pollack79,cuk04}, although
the disc does not play such a role  in the Nice model
 \citep[e.g.][]{gomes05}.  The circumplanetary accretion disc
plays a vital role in satellite formation. The  
regular satellite systems of Jupiter and Saturn extend
over a small fraction of their respective Hill radii, to less than $0.06 r_{\rm H}$. 
This fact has motivated a dynamical explanation in terms of 
circumplanetary disc sizes \citep[e.g.][]{canup02, mosqueira03}. We discuss this in more
detail in Section 8.

Previous studies have considered the role of the angular momentum
(about the planet) of the inflowing gas in determining the disc
structure, in particular the disc outer radius.  This picture has some
intuitive appeal, since no disc would form, i.e., the disc radius
would be zero, if the inflowing gas has zero angular momentum.
\cite{quillen98} suggested that after the planet has opened a gap, the
circumplanetary disc extends to $r \approx r_{\rm H}/3$ . 
This result
is based on the initial angular momentum of gas whose flow relative to
the planet is slow near the planet's Hill sphere and subsequently
accelerates inward towards the planet while conserving its angular
momentum about the planet.    \cite{canup02} and \cite{ward10}  considered cases
where the angular momentum of the inflowing gas is considerably smaller. 

\cite{lissauer95}  (see also  \citealt{lissauer09}) suggested that the
circumplanetary disc structure prior to or during gap opening could account for
the location of regular satellites. This model was also based on the
concept that the disc structure is determined by the angular momentum
of infalling gas about the planet, with an implied smaller value of disc radius of $r=
r_{\rm H}/48$. 
This paper concentrates on the former case, the case
of a circumplanetary disc in the presence of gap opening.
The results of this paper may also apply to this case.
We briefly discuss this point again below equation (\ref{Hr}) and
in Section \ref{concs}.

This radius characteristic of the angular momentum of the accreting
gas is useful for providing an estimate of the conditions required for
circumplanetary disc formation.  For a disc to form about a planet
with a gap, the planetary radius must be smaller than this value.
Violating this constraint for disc formation would require very short
orbital periods, of order $1.5\,\rm d$, for a planet of Jupiter's
radius.

However, it is not clear that this radius characteristic of the
angular momentum of the accreting gas is important for determining the
disc radius or even has much influence of the disc structure.  The
reason has to do with the requirements of angular momentum loss from a
steady-state disc.  The process of angular momentum transport in
circumplanetary discs has up to now been largely ignored. 
 The work 
 by \cite{canup02} and \cite{ward10} did consider the angular momentum issue, 
 as will be discussed further in Section 6.5. \cite{mosqueira03} briefly discuss the possible
 role of resonances in truncating the disc. For the disc to
accrete in a steady-state manner, gas within it must continuously lose
angular momentum.  These studies recognized that the disc would
 extend beyond the radius in which mass is injected in order to account for the requirements
 of angular momentum conservation. The angular momentum issue has been previously encountered in the
case of discs in mass exchange binary star systems
\citep[e.g.][]{papaloizou77}. The inflowing gas from the mass losing
star is captured within the Roche lobe of the companion star.  This
gas has some angular momentum that often results in the formation of
disc about the companion. The picture that has resulted is that the
disc spreads outward by viscous torques until it reaches a radius at
which tidal torques from the companion are sufficient to remove the
angular momentum at the rate required for steady state accretion.  The
disc structure is insensitive to the angular momentum of the accreting
gas. The case of a planet accreting gas can be considered to be a
binary of extreme mass ratio. The accretion disc model in the binary
star case suggests a somewhat different picture of how the accretion
process operates than has been previously considered for
circumplanetary discs. We consider the accretion disc model of
circumplanetary discs in this paper.

Some recent three-dimensional simulations have analysed the flow in
circumplanetary discs.  For example, \cite{ayliffe09} simulated gas
accretion by protoplanets in three-dimensions with gas self-gravity
and radiation transfer to investigate the properties of
circumplanetary discs. They determined the disc scale height to be
$H/r>0.2$ and found the disc size to be about $0.35 r_{\rm H}$.
They attributed this value of the disc radius as following from the estimate
of \cite{quillen98}, for which  the radius is determined by the angular
momentum of the accreting gas. 

\cite{machida09} also performed three-dimensional calculations of
circumplanetary discs.  For a Jupiter mass planet, he found a peak in
the surface density at about 21 protoplanetary radii (or
$0.028\,r_{\rm H}$) from the protoplanet centre. According to these
results, the peak is due to the balance between the centrifugal force
and the gravity of the protoplanet.  Such small scale features might
then correspond to the  formation sites of the regular satellites
around Jupiter and Saturn. However, this peak in the surface density
was not reproduced by \cite{ayliffe09} who found that the surface
density decreases monotonically with radius from the protoplanet.

We analyse the general properties of circumplanetary discs by the use
of some simplified models, in order to understand the physical
processes that affect disc structure. In Section~\ref{properties}, we
estimate the scale height and temperature of the disc. In
Section~\ref{test} we consider a ballistic disc model consisting of
periodic particle orbits around a planet. In Section~\ref{eqs} we
examine the circumplanetary fluid disc evolution equations in the Hill
approximation and discuss their relationship to the binary star case.
In Section~\ref{res} we consider whether resonances could lie within a
circumplanetary disc. In Section~\ref{tidal}, we analyse the effects
of a tidal torque on a purely viscous disc.  In Section~\ref{sph} we
describe results of a model that includes the effects of gas pressure
by means of some SPH simulations.  Section~\ref{concs} contains a
discussion and the conclusions.

\section{Circumplanetary Disc Properties}
\label{properties}

We consider a system with a planet of mass $M_{\rm p}$ in a circular
orbit around a star of mass $M_{\rm s}$ at a separation $a$.  The Hill
sphere is the approximate region where the planet's gravity dominates
that of the star and its radius is given by
\begin{equation}
r_{\rm H}=a\left(\frac{\mu}{3}\right)^\frac{1}{3},
\label{rh}
\end{equation} 
 where $\mu=M_{\rm p}/(M_{\rm p}+M_{\rm s}) \ll 1$.  
 The sound speed in an ideal gas is
\begin{equation}
c_{\rm s}= \sqrt{\frac{kT}{\mu_{\rm m} m_{\rm H}}} \approx 10^6 \sqrt{\frac{T}{10^4}} \,\rm cm\, s^{-1},
\end{equation}
where $k$ is the Boltzmann constant, $T$ is the temperature, $\mu_{\rm
  m}$ is the mean molecular weight and $m_{\rm H}$ is the mass of a
hydrogen atom. The surface temperature of a steady state accretion
disc around an object of mass $M$ at a radius $r$ is given by
\begin{equation}
\sigma T^4= \frac{3}{8\pi}\frac{GM \dot M}{r^3}
\label{temp}
\end{equation}
\citep{pringle81} where $\sigma$ is the Stefan-Boltzmann constant.

We estimate the temperature of the accreting gas with the disc.
Similar approaches were taken by \cite{canup02}.  Consider a disc
orbiting a solar mass star that is accreting gas at a rate of $10^{-8}\,\rm
M_\odot \, yr^{-1}$, a typical value inferred from observations of T
Tauri stars.  Equation~\ref{temp} implies that at a distance from the
central star of $5\,\rm AU,$ the disc has a temperature of about
$26\,\rm K$, ignoring effects of stellar heating.  This temperature
corresponds to a disc aspect ratio $H/r \simeq 0.04$.

The aspect ratio of the circumplanetary disc taken with respect to the planet is
\begin{equation}
\left(\frac{H}{r}\right)_{\rm p}= \frac{c_{\rm s}}{\Omega \, r}= 0.3
\left(\frac{r}{r_{\rm H}/3}\right)^{\frac{1}{8}} \left(\frac{M_s}{M_\odot} \right)^{\frac{1}{24}}
\left(\frac{M_{\rm p}}{M_{\rm J}}\right)^{-\frac{1}{3}}
\left(\frac{a}{5\,\rm AU}\right)^\frac{1}{8}
\left(\frac{\dot M_{\rm p}}{10^{-8}\rm \, M_\odot\,yr^{-1}}\right)^\frac{1}{8},
\label{Hr}
\end{equation}
where $M_{\rm J}$ is the mass of Jupiter and $r$ is the distance from
the planet. (There is a small correction due to the vertical gravity
of the star that we ignore.) This aspect ratio is much larger than the
value for the circumstellar disc, $(H/r)_{\rm s} \simeq 0.04$. We have
taken the disc sound speed $c_{\rm s}$ to be given by its value at the
disc surface and have ignored any increase at the midplane with
optical depth.  The relatively high value of $(H/r)_{\rm p}$ is due to
the higher gas temperatures and the typically smaller $\Omega r$
values in the circumplanetary disc case.  The value of the disc aspect
ratio from equation (\ref{Hr}) is similar to the values obtained in
the simulations by \cite{ayliffe09}.  They found no evidence
for circumplanetary discs in their simulations of low mass planets (less than
100 Earth masses).  They attributed this result to the weakness of the planet's
vertical gravity compared to pressure, as  can be seen in equation (\ref{Hr}).
For a sufficiently low mass planet, $(H/r)_p$ can be large enough that the concept
of a disc breaks down and the gas around the planet is essentially spherical. 
As is well known, the
circumplanetary disc temperatures at these accretion rates are too
high to explain the existence of the icy satellites of Jupiter and
Saturn \citep{canup02}.  However, in a slower accretion phase as
proposed by \cite{canup02} and further investigated by \cite{barr08},
the temperature would be lower and $H/r$ would be smaller, $H/r \sim
0.1$.  We discuss this further in Section~\ref{concs}.

We compare the surface density in the circumplanetary disc,
$\Sigma_{\rm p}$, to the surface density in the circumstellar disc,
$\Sigma_{\rm s}$. We assume that most of the mass being accreted
through the circumstellar disc outside the orbit of the planet is
accreted by it.  As discussed in the Introduction, this efficient gas
capture has been found in simulations of Jupiter mass planets that
orbit solar mass stars.  The mass transfer rate through a steady state
disc is given by $\dot M \propto \nu \Sigma$ \citep{pringle81}, where
the viscosity is parametrised with the $\alpha$ prescription so that
\begin{equation}
\nu=\alpha c_{\rm s}H = \alpha \left(\frac{H}{r}\right)^2 r^2\Omega
\label{alphavisc}
\end{equation}
\citep{shakura73} where $\Omega=\sqrt{GM_{\rm p}/r}$.  It then follows
that
\begin{equation}
\frac{\Sigma_{\rm s}}{\Sigma_{\rm p}}\sim \frac{\nu_{\rm p}}{\nu_{\rm s}} \sim  
\mu^{2/3} \frac{\alpha_{\rm p}}{\alpha_{\rm s}}    \frac{ \left(\frac{H}{r}\right)^2_{\rm p}}{\left(\frac{H}{r}\right)^2_{\rm s}} 
  \sqrt{ \frac{r}{r_{\rm H}} },
\end{equation}
where  $\nu_{\rm s}$ and $\nu_{\rm p}$ are the viscosities in the
circumstellar and circumplanetary discs respectively, and $\alpha_{\rm s}$
and $\alpha_{\rm p}$ similarly refer the $\alpha$ values in the discs. 

Consider a solar mass star and a Jupiter mass planet, $\mu=10^{-3}$,
and circumstellar and circumplanetary disc aspect ratios of $0.04$
and 0.3 respectively, as discussed above. For $r= 0.2 r_{\rm H}$ 
and equal $\alpha$ values ($\alpha_{\rm s} = \alpha_{\rm p}$)
in the above equation, the ratio of the surface densities is
$\Sigma_{\rm s}/\Sigma_{\rm p} \sim 0.25$.  So the circumplanetary
disc surface density is somewhat higher that the local circumstellar
disc density. Circumstellar discs could under some circumstances
contain dead zones, regions where the disc ionization is too low for
the magneto-rotational instability to operate as a source of
turbulence (Gammie 1996).  Dead zones operate where the temperature is
sufficiently low that thermal ionization is weak (less than about
$10^3$ degrees) and where the surface densities are high enough that
external sources of ionization such as cosmic rays do not penetrate
far below the disc surface.  For expected conditions in a
circumstellar disc, dead zones may extend from a few tenths of an AU
to several AU, in some cases beyond Jupiter's orbital radius
\citep[e.g.][]{terquem08}. The model here suggests that the conditions
for dead zone formation within a circumplanetary disc are perhaps more
favorable than in the nearby circumstellar disc gas. The reason is
that the circumstellar disc surface densities are higher than in the
nearby circumstellar disc gas, while the temperatures are low enough
to avoid sufficient ionisation in much of the circumplanetary disc. A
difference in the properties of turbulence in the two discs violates
our assumption of $\alpha_{\rm s} = \alpha_{\rm p}$. But, it does so in a
way that may further enforce the importance of dead zones in the
circumplanetary disc. That is, $\alpha_{\rm s} / \alpha_{\rm p}$ would be
expected to increase, leading to a decrease in $\Sigma_{\rm
  s}/\Sigma_{\rm p}.$ We do not pursue the possibility of dead zones
further in our disc models described later and assume a simple viscous
disc. 
  
The viscous timescale in the disc is
\begin{equation}
t_{\nu_{\rm p}} \sim \frac{r^2}{\nu_{\rm p}} \sim \frac{1}{\alpha (H/r)_{\rm p}^2 \Omega} \sim 
10^3 P \left( \frac{10^{-3}}{\alpha} \right) \left(\frac{0.3}{H/r} \right)^2 \left(\frac{r}{r_{\rm H}} \right)^{3/2},
\label{tauvisc}
\end{equation}
where $P$ is the orbital period of the planet. So we expect that such
discs should be viscously relaxed for Jupiter and Saturn. This result
has the implication that the viscous disc flow can be regarded to be
in a steady state.

 A crude estimate of the steady state luminosity ratio of the
 circumplanetary to circumstellar discs in a system with a Jupiter mass planet is
\begin{equation}
\frac{L_{\rm p}}{L_{\rm s}} \sim \frac{\dot{M}_{\rm p} M_{\rm p} R_{\rm s}}
{\dot{M}_{\rm s} M_{\rm s} R_{\rm p}},
\label{lum}
\end{equation}
where $R_{\rm p}$ and $R_{\rm s}$ are the planet and star radii,
respectively.  For $\dot{M}_{\rm p} \sim 1-10 \dot{M}_{\rm s}, M_{\rm
  p} \sim 0.001 M_{\rm s},$ and $R_{\rm p} \sim 0.1 R_{\rm s}$, we
estimate $L_{\rm p}/ L_{\rm s}$ to be of order a few percent.  We see
that circumplanetary discs are not very bright. We discuss this
further in Section~\ref{concs}.

\section{Ballistic Particle Periodic Orbits Around the Planet}
\label{test}

In this section we consider ballistic particle orbits in a planet-star
system in order to find their nonKeplerian angular velocity and where
the orbits begin to cross.

\subsection{Ballistic Equations}
As a simple two-dimensional disc model, we consider ballistic
particles orbiting a planet of mass $M_{\rm p}$ in the corotating
frame of the star-planet system with a star of mass $M_{\rm s}$. The
particles lie in the star-planet orbit plane. For such a model to
represent a low pressure (cold) steady state fluid disc, each orbit
must be periodic in the corotating frame and be nonintersecting,
either with itself or neighboring orbits, in order that the velocity
be single-valued in space. In addition, the orbit must be stable. The
equation of motion of a ballistic particle at position $\bm{r}$ for
potential $\phi$ in the corotating frame is
\begin{equation}
\bm{\ddot r}+ 2 \bm{\Omega}_{\rm p} \times \bm{\dot r}=-\bm{\nabla \phi} ,
\label{eqmot}
\end{equation}
where $\bm{\Omega}_{\rm p}$ is the angular velocity of the planet. The
potential due to the rotation of the frame and the gravity of the
planet and the star is given by
\begin{equation}
\phi= -\frac{|\bm{r}- \bm{r}_{\rm cm}|^2 \, \Omega_{\rm p}^2}{2}-\frac{G M_{\rm p}}{|\bm{r}-\bm{r}_{\rm p}|}-\frac{G M_{\rm s}}{|\bm{r}-\bm{r}_{\rm s}|} ,
\label{phif}
\end{equation}
where $\bm{r}_{\rm cm}$, $\bm{r}_{\rm p}$ and $\bm{r}_{\rm s}$ are the
position vectors of center of mass, the planet, and the star
respectively.  The angular velocity of the star--planet binary system
is
\begin{equation}
\Omega_{\rm p}=\left(\frac{G (M_{\rm s}+M_{\rm p})}{a^3}\right)^\frac{1}{2}.
\label{omegap}
\end{equation}

We adopt units where $a=M_{\rm s}+M_{\rm p}=\Omega_{\rm p}=1$. The
equations of motion for the particle with Cartesian coordinates
$(x,y)$, whose origin is at the center of mass, are then
\begin{equation}
\ddot x - 2 \dot y -  x + \frac{\mu \left[ x-(1-\mu)\right]}{\left(\left[x-(1-\mu)\right]^2+y^2\right)^\frac{3}{2}} +\frac{(1-\mu)(x+\mu)}{\left[(x+\mu)^2+y^2\right]^\frac{3}{2}}=0
\label{xeqtest}
\end{equation}
and
\begin{equation}
\ddot y + 2 \dot x -  y + \frac{\mu y}{\left(\left[x-(1-\mu)\right]^2+y^2\right)^\frac{3}{2}} +\frac{(1-\mu)y}{\left[(x+\mu)^2+y^2\right]^\frac{3}{2}}=0.
\label{yeqtest}
\end{equation}

Following the standard procedure to obtain the Hill equations, we
transform these equations to a coordinate system centered on the
planet and rescale the coordinates.  The scaling is chosen so that the
new coordinates scale with the Hill radius, see equation (\ref{rh}).
We make the change of variables to rescaled radius $R$ defined by
\begin{equation}
\bm{R}=(\bm{r}-\bm{r}_{\rm p-cm})/(a \mu^{\frac{1}{3}}),
\label{Rscaled}
\end{equation}
where $\bm{r}_{\rm p-cm}$ is the displacement of the planet from the
center of mass.  The equation of motion (\ref{eqmot}) is then
\begin{equation}
\bm{\ddot R}+ 2  \bm{\hat{e}}_{z} \times \bm{\dot R}=-\bm{\nabla} \Phi,
\label{neweqmot}
\end{equation}
where $\Phi$ is the potential in this new frame that we determine in
Section~\ref{potential}. We let $X = 1-\mu +\mu^\frac{1}{3}x$ and $Y
=\mu^\frac{1}{3}y$. Since $\mu$ is small, we consider here only the
terms to lowest order in $\mu$. The equations to order
$\mu^\frac{1}{3}$ become
\begin{equation}
-3X + \frac{X}{(X^2+Y^2)^\frac{3}{2}}-2 \dot Y +\ddot X=0
\label{xeq}
\end{equation}
and 
\begin{equation}
\frac{Y}{(X^2+Y^2)^\frac{3}{2}}+2\dot X+\ddot Y=0.
\label{yeq}
\end{equation}
In Section~\ref{cross} we consider the higher order terms in $\mu$
that have been neglected here.

\subsection{The Potential in the Hill Approximation}
\label{potential}

The terms in equations~(\ref{xeq}) and~(\ref{yeq}) that are functions
of $X$ and $Y$ are expressed as the potential gradients
\begin{equation}
\frac{\partial \Phi}{\partial X}=-3X+\frac{X}{(X^2+Y^2)^\frac{3}{2}}
\end{equation}
and
\begin{equation}
\frac{\partial \Phi}{\partial Y}=\frac{Y}{(X^2+Y^2)^\frac{3}{2}}.
\end{equation}
Integrating these, we find the potential in the Hill approximation to
be
\begin{equation}
\Phi=-\frac{1}{(X^2+Y^2)^\frac{1}{2}}-\frac{3}{2}X^2.
\end{equation}
In polar coordinates centered on the planet, so that $X=R \cos \theta$
and $Y=R\sin \theta$, the potential is given by
\begin{align}
\Phi & =-\frac{1}{R}-\frac{3}{2} R^2 \cos^2\theta \cr
& = -\frac{1}{R}- \frac{3}{4}R^2-\frac{3}{4}R^2 \cos 2 \theta.
\label{phi}
\end{align}
The first term is a point mass potential of the planet. The other two
terms are due to the rotation of the frame and the gravitational
effects of the star.

In dimensional form, the potential is given by
\begin{align}
\phi_{\rm H} & =- \Omega_{\rm p}^2 a ^2 \mu \left (\frac{a}{r}+\frac{3}{4 \mu}  \left(\frac{r}{a}\right)^2 (1+ \cos 2 \theta)
\right),
\label{phiH}
\end{align}
which is valid for $r \sim O (\mu^{1/3} a)$.

\subsection{Angular Velocity in the Hill Approximation}
\label{ang}

From the $r$-component of the equation of motion~(\ref{eqmot}) with $\phi=\phi_{\rm H}$, we
have that
\begin{equation}
-\frac{u_\theta^2}{r}-2 \Omega_{\rm p} r (\Omega - \Omega_{\rm p}) = -\frac{\partial \phi_{\rm H}}{\partial r}.
\end{equation}
We substitute the axisymmetric terms in the potential of equation~(\ref{phiH}) and the
velocity in the $\theta$-direction, $u_\theta=r(\Omega  - \Omega_{\rm p})$, to find
\begin{equation}
\Omega^2 = \frac{GM_{\rm p}}{r^3}-\frac{1}{2}\Omega_{\rm p}^2,
\label{angvel}
\end{equation}
where $r$ is the (unscaled) distance from the planet.  There are
corrections to this result due to nonlinear effects of nonaxisymmetric
terms.  The angular velocity about the planet, $\Omega$, is less than
Keplerian because of the effects of the star.

\subsection{Crossing and Unstable Orbits}
\label{cross}

\begin{figure}
\centering
\includegraphics[width=6.4cm]{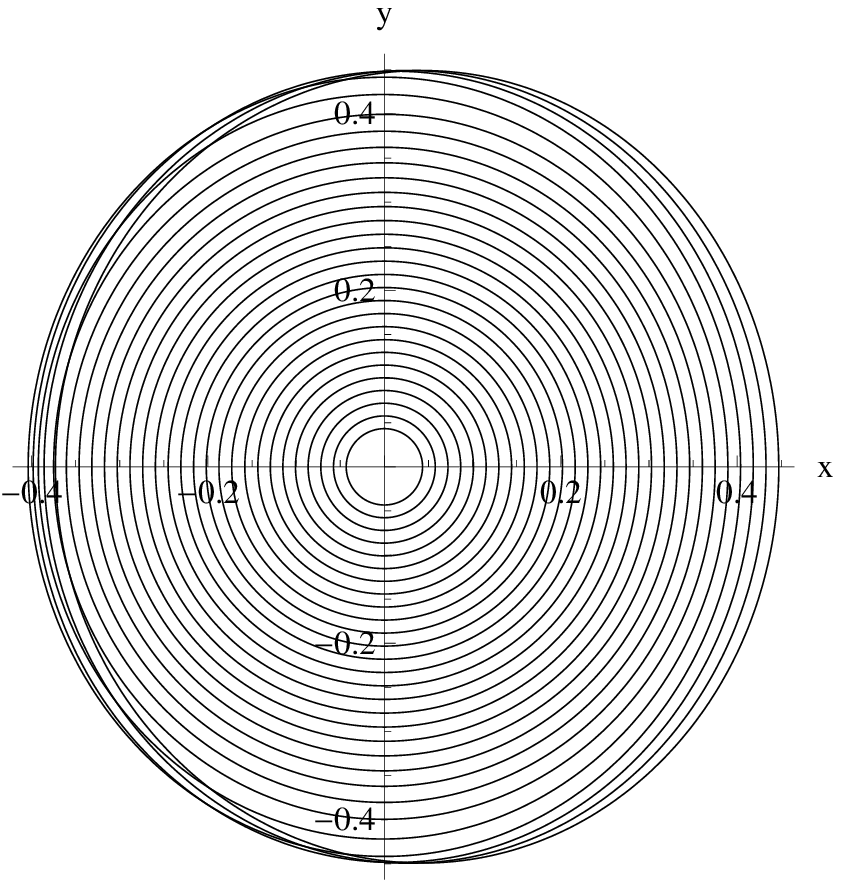}
\includegraphics[width=6.4cm]{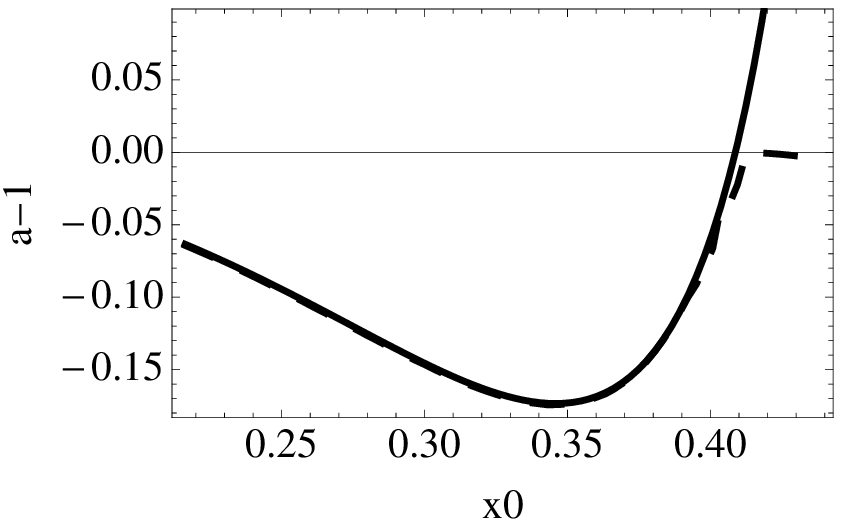}
\caption{Left: Periodic orbits about a planet in the Hill
  approximation. The $x$ and $y$ coordinates are in units of the Hill
  radius $r_{\rm H}$. At a distance of about $0.4 r_{\rm H}$ from the
  planet, the orbits begin to cross. Right: Stability parameter (Henon
  1965) $a-1$ plotted against $x_0$, the $x$ coordinate  in units of
  $r_{\rm H}$ where the orbit crosses the positive $x$-axis. For $a-1$
  negative (positive) the orbits are dynamically stable (unstable). The
  dashed line is for periodic orbits that eventually intersect each other at
  larger $x_0$, as occurs in the left panel.  The solid line tracks
  the stability of periodic orbits that remain symmetric about the
  $y$-axis for larger $x_0$.  Notice that for $x_0 \simeq 0.4 r_{\rm H}$, where
  orbits begin to cross in the left panel, there is a bifurcation in
  stability properties.  At larger radii, $y$-symmetric orbits remain
  nonintersecting, but are unstable, while the intersecting orbits are
  marginally stable.  }
\label{orbitcross}
\end{figure}

We numerically solved equations~(\ref{xeq}) and~(\ref{yeq}) to find
the trajectories of particles in this potential. We found closed
periodic particle orbits in the frame of the planet, along the lines
of \cite{paczynski77}, but in the Hill approximation.  We chose the
initial position for the particle to be on the $x$-axis. The initial
$y$-velocity is determined so that the particle has zero velocity
in the $x$ direction when it again crosses the $x$-axis on the other
side of the central planet.  We find that the neighboring particle
orbits first intersect for the initial position on the $x$-axis of $x=0.410 r_{\rm H}$ (or $X=0.294$), as shown in
the left panel of Fig.~\ref{orbitcross}.  This maximum radius of nonintersecting orbits, $r_{\rm max} \simeq
0.410 r_{\rm H}$,
is considered to represent a maximum disc radius for a cold disc, as
discussed by \cite{paczynski77}.  

The orbits in in the left panel Fig.~\ref{orbitcross} are asymmetric
(lopsided) with respect to the y-axis in that the intersections occur
for $x<0$. On the other hand, the $m=2$ Hill potential is symmetric
about the y-axis.  A more detailed analysis shows that there is
another set of orbits for which the intersections occur for $x>0$. In
addition, there is a third set of orbits that remain symmetric about
the y-axis and nonintersecting for $r > r_{\rm max}$. However, this
set of orbits is dynamically unstable for $r > r_{\rm max}$. That is,
there is a bifurcation that occurs at $r = r_{\rm max}$, as seen in
the right panel of Fig.~\ref{orbitcross}. Beyond this radius, there
continue to exist y-symmetric nonintersecting orbits, but they are
unstable.  There are stable orbits beyond this radius, but they
intersect with neighboring orbits. In any case, this critical radius of $r_{\rm
  max}=0.410\, r_{\rm H}$ is the limiting radius for orbits that
represent a cold, steady-state disc.

To test the validity the Hill approximation, we also solved the
equations of motion for the full gravitational potential in the frame
of the planet. We determined the particle orbits in the same way as
described above. We found that the orbits first intersect at a radius
of $0.412\, r_{\rm H}$ with a mass ratio $\mu=0.01$. The Hill
approximation is then reasonably accurate for these purposes, even for
relatively large mass planets.

\subsection{Higher Order terms}

By  including terms of O($\mu^\frac{2}{3}$) and O($\mu$) in
equations~(\ref{xeq}) and~(\ref{yeq}), we obtain the dimensionless potential 
\begin{equation}
\Phi  = -\frac{1}{R}- \frac{3}{4}R^2-\frac{3}{4}R^2 \cos 2 \theta
+\frac{\mu^\frac{1}{3}}{8} R^3  \left(3 \cos\theta - 5 \cos3\theta\right)
-\frac{\mu^\frac{2}{3}}{64} R^4 \left(9 + 20 \cos2\theta + 35 \cos4\theta\right).
\label{phiex}
\end{equation}
These higher order terms in $\mu^{1/3}$ introduce additional angular
dependencies.

\section{Disc Fluid Equations}
\label{eqs}

The steady-state two-dimensional equations of motion for a fluid disc in the corotating
frame are
\begin{equation}
 \bm{\nabla} \cdot (\Sigma {\bf u}) =0
\label{origa}  
\end{equation} 
and 
\begin{equation}
(\bm{u \cdot \nabla})\bm{u} + 2 \bm{\Omega}_{\rm p}
\times \bm{u} = -\bm{\nabla \phi}-\frac{1}{\Sigma}\bm{\nabla} p +\bm{f}_{\rm v} ,
\label{orig}
\end{equation}
where $\bm{u}$ is the fluid velocity and $\bm{f}_{\rm v}$ is the
 force per unit disc mass that
represents the effects of turbulent viscosity. Quantity $p$ is the
two-dimensional (vertically integrated) pressure.  We consider a
cylindrical coordinate system $(r, \theta)$ centered on the planet.
The azimuthal viscous force due to shear is given by
\begin{equation}
f_{r,\theta}= \frac{1}{r^2\Sigma} \frac{\partial}{\partial r}\left( \nu \Sigma r^3 \frac{d\Omega}{dr}\right),
\end{equation}
where $\Sigma$ is the surface density and $\nu$ is the kinematic
turbulent viscosity.  The potential $\phi$ is given by equation
(\ref{phif}).

We rescale the variables by $\mu^{1/3}$ as we did for the ballistic
case in equation~(\ref{Rscaled}). We rescale the velocity and radius
as $\bm{U}=\bm{u}/(\mu^\frac{1}{3} a  \Omega_{\rm p} )$ and $R = r/(\mu^{1/3} a)$,
respectively, and obtain
\begin{equation}
\bm{ \tilde{\nabla}} \cdot (\Sigma {\bf U}) =0
\label{origas}
\end{equation}
and
\begin{equation}
 (\bm{U \cdot \tilde{\nabla}}) \bm{U} +
2\bm{\tilde{\Omega}}_{\rm p} \times \bm{U} = -\bm{ \tilde{\nabla}}
\Phi-\frac{1}{\Sigma}\bm{ \tilde{\nabla}} P + \bm{F}_{\rm v},
\label{origs}
\end{equation}
where
\begin{equation}       
\bm{F}_{\rm v} =  \bm{f}_{\rm v} / (\mu^{1/3} a^2  \Omega_{\rm p} ), 
\end{equation}
\begin{equation}
\bm{ \tilde{\nabla}} =  (\mu^{1/3} a)  \bm{\nabla},
\end{equation}
\begin{equation}
P =  p/(\mu^{2/3}  a^2 \Omega_{\rm p}^2),
\end{equation}
\begin{equation}
\bm{\tilde{\Omega}} =  \bm{\Omega}/\Omega_{\rm p},
\end{equation}
and
\begin{equation}
\Phi = \phi/(\mu^{2/3} a^2 \Omega_{\rm p}^2).
\end{equation}
The potential, $\Phi$, is given by equation (\ref{phiex}) and is
independent of $\mu$ to lowest order, the Hill approximation.  The
rescaled viscous shear force is
\begin{equation}
F_{r,\theta} = \frac{1}{R^2 \Sigma}\frac{\partial}{\partial R}\left( \alpha R^5 \Sigma \left(\frac{H}{r}\right)^2 \tilde{\Omega} \frac{d \tilde{\Omega}}{dR}\right),
\end{equation}
where we used the $\alpha$-prescription for the viscosity so that
\begin{equation}
\nu = \alpha \left(\frac{H}{r}\right)^2 r^2 \Omega = \mu^{2/3}  \alpha \left(\frac{H}{r}\right)^2 R^2 \tilde{\Omega} \, a^2 \Omega_{\rm p}
\end{equation}
\citep{shakura73}. We have left the disc aspect ratio about the planet
in the unscaled form $H/r$ because it is invariant to scaling. We have not
rescaled the surface density because doing so would have no effect on the
velocity field, since we neglect disc self-gravity. Notice
that $F_{r,\theta}$ is independent of the mass ratio $\mu$.  This is
also true of all other components of the viscous force.  The rescaled
pressure is given by
\begin{align}
 P  = \Sigma \left(\frac{H}{r}\right)^2 R^2  \tilde{\Omega}^2,
\end{align}
which is independent of $\mu$.  Thus, the transformed equations
(\ref{origas}) and (\ref{origs}) are of the same form as the original
equations (\ref{origa}) and (\ref{orig}) and are independent of $\mu$
in the Hill approximation. This result also holds in an obvious extension
to three dimensions.

The rescaled circumplanetary disc equations are then very similar to
the circumstellar disc equations previously analysed for close binary
stars with order unity mass ratio. One difference is the form of the
potential, which has a single nonaxisymmetric azimuthal number $m=2$
in the Hill approximation.  However, this is the dominant tidal term
in close binaries.  Consider a disc in binary that orbits star 1 and
is tidally perturbed by star 2.  At a small distance $r \ll a$ from
star 1, the potential terms due to the star 1 and the $m=2$ tidal
component due to star 2 is given by $\phi = -G M_1/r -3 G M_2 \,
\cos(2\theta) \, r^2/(4a^3)$. This potential is similar (within a
factor of 2) to the rescaled potential of equation (\ref{phi}) with
$G=1, a=1$, and masses $M_1 = M_2= 1/2$.  In particular, the ratio of
the $m=2$ tidal to central potentials is the same for both cases.
Therefore, the disc flow equations for a planet in the Hill
approximation are similar to those for a binary star system with unity
mass ratio, for a given value of $H/r$ and $\alpha$.

\section{Resonances}
\label{res}

Resonances occur in the disc where the forcing frequency matches a
natural frequency in the disc. Angular momentum can be transferred to
the star from the disc by the tidal torques that are exerted at the
Lindblad resonances within the disc \citep{goldreich79}.  At such resonances, rotationally
modified pressure waves are launched.  Torques are exerted on the disc
at radii where the waves damp.  Resonance torques could play a role in
truncating the disc. In this section we consider whether circular,
eccentric, or vertical resonances could lie within the disc. We only
consider nearly circular orbits in this analysis. We find this
approximation holds well out to the radius where orbits begin to cross
(Section~\ref{cross}).  Since we expect the disc to be truncated
inside or at this radius, we should find all of the resonances present
in the disc in this approximation.

However, for a disc as warm as suggested during the T Tauri accretion
phase, $H/r \sim 0.3$ (equation (\ref{Hr})), some gas might extend beyond the orbit crossing radius
and off-resonant forcing could
play a role. 
Off-resonant forcing is possible because for $m$ of order unity, the resonance
width scales as $(H/r)^{2/3}\, r_{\rm H}$.  With such a large width,
the resonance could overlap with the disc, even though the exact
resonance location does not lie within the main body of the disc, as has been investigated in
the binary star case by \cite*{savonije94}.

\subsection{Circular Resonances}

We apply the angular velocity of the disc in the Hill approximation
given in equation~(\ref{angvel}) to the Lindblad
resonance condition \cite{goldreich79}. The circular Lindblad resonances
then occur where
\begin{equation}
-m^2 (\Omega-\Omega_{\rm p})^2 +4 \Omega \left(\Omega +\frac{r}{2} \frac{d \Omega}{dr}\right)=0.
\end{equation}
In the Hill approximation we have only the $m=2$ term of the
potential. We find that the only positions inside the orbit crossing
radius is for the $m=1$ term which is at $R=0$.  For a cold disc, the
resonance width is very small and the tidal forcing is very weak near
the disc center. For a warm disc, stronger resonant excitation is
possible.  At higher order in $\mu^{1/3}$, other $m$-values are
present (equation~\ref{phiex}). However, these higher order resonances
also fail to lie inside the orbit crossing radius.

\subsection{Eccentric Resonances}

If the orbit of the planet is slightly eccentric with eccentricity
$e$, the tidal forcing can be decomposed into a series or rigidly
rotating potentials at various frequencies $\ell \, \Omega_{\rm p}$
for integer $\ell$ \citep{goldreich79}.  The eccentric Lindblad
resonances then occur where
\begin{equation}
-(m \Omega  -\ell \Omega_{\rm p})^2 +4 \Omega \left(\Omega+\frac{r}{2} \frac{d \Omega}{d r}\right)=0.
\end{equation}
For $m=2$, the lowest order resonance that lies within the orbit
crossing radius has $\ell =7$.  The resonant torque scales as $e^{2
  |l-m|}$, where $m$ is the azimuthal wavenumber of the tidal
potential. Consequently, the torque would scale as $e^{10}$, which
suggests that the torque would be quite weak for modest
eccentricities.  For such a resonance to be able to overcome the
effects of disc turbulent viscosity and truncate the disc, we roughly
require $\alpha \la (r/H)^2 \,e^{10}$ \citep[see,][]{artymowicz94}.
Consequently, the disc viscosity would need to be very small.

\subsection{Vertical Resonances}

There is also a set of resonances associated with the vertical disc
motions. For each $m$, the vertical disc resonance lies closer to the
planet than the corresponding horizontal (coplanar) Lindblad resonance, although
the torque it produces is a factor of $(H/r)^2$ smaller than the
corresponding horizontal resonance. The vertical resonance generates
horizontally propagating waves which travel towards the planet at the
centre of the disc. The wave generation can transport angular momentum
to the orbital motion of the planet-star system.  The vertical
resonances occur where
\begin{equation}
m^2(\Omega-\Omega_{\rm p})^2=(1+\Gamma)\left(\Omega_{\rm p}^2+\frac{G M_p}{r^3}\right),
\end{equation} 
where $\Gamma$ is the effective adiabatic index \citep{lubow81}.  We
solve this equation with $\Gamma=1.4$ and find the $m=2$ resonance
occurs at $r/r_{\rm H}=0.49$, which is somewhat outside the orbit
crossing radius.  Resonances at higher $m$-values occur further away
from the planet. The analysis in subsequent sections of this paper considers only one- and two-dimensional
models that are not capable of finding vertical resonances.

\section{Viscous Disc Model}
\label{tidal}

We consider the effect of the tidal perturbation on a circumplanetary
accretion disc due to the presence of the star by following the work
of \cite{papaloizou77}.  The tidal torque transfers angular momentum
from the disc to the binary orbit, allowing material to accrete on to
the central object \citep{borner73,lin76}.  The disc attempts to
expand by viscous forces, but tidal torques dominate the transport
process only in the outer parts of the disc.  In this approach, the
disc pressure forces are ignored. Consequently, this approach does not
include the effects of off-resonant waves that could be present.  It
is more accurate for cool discs, $H/r < 0.1$, where the effects of
waves are likely less important. Such cool circumplanetary discs could
arise at late stages of disc evolution where the accretion rates are
lower as the disc disperses.

\subsection{Linearised Velocity}

We consider linearised equations for the tidal disturbances about the
circular motions around the planet and initially include only
gravitational and centrifugal forces. The unperturbed gas orbits with
velocity $\bm{u_0}=(0,r (\Omega-\Omega_{\rm p}))$ in the corotating
frame centered on the planet, where $\Omega$ is given by equation
(\ref{angvel}). We consider perturbations to this flow by the presence
of the star.  We denote the perturbed velocity as $\bm{u_1}$, where
$u_1 \ll u_0$ and $\bm{u_1}=(u_r,u_{\theta})$.  We follow the approach
of \cite{papaloizou77}, but adopt the corotating frame with the Hill
approximation.  The linearised equations of motion based on
equation~(\ref{orig}) are
\begin{equation}
(\Omega-\Omega_{\rm p})\frac{\partial u_r}{\partial \theta}-2 \Omega u_\theta 
=f_r
\end{equation}
and
\begin{equation}
(\Omega-\Omega_{\rm p})\frac{\partial u_\theta}{\partial \theta}+2 B u_r
=f_\theta,
\end{equation}
where
\begin{equation}
B=\frac{r}{2}\frac{d\Omega}{dr}+\Omega.
\end{equation}
The force components are obtained from the negative gradient of
potential $\phi_{\rm H}$ in equation~(\ref{phiH}).

Since only $m=2$ disturbances are involved in the Hill approximation, we represent
\begin{equation}
u_r, u_\theta, f_R, f_\theta \propto  \exp(2 i \theta)
\end{equation}
and obtain
\begin{equation}
u_r= \frac{ i  (\Omega -\Omega_{\rm p}) f_r  +   \Omega f_\theta}
{2 [ - (\Omega-\Omega_{\rm p})^2 +  B  \Omega  ]}
\label{ur}
\end{equation}
and
\begin{equation}
u_\theta =  \frac{   - B  f_r  + i   (\Omega -\Omega_{\rm p}) f_\theta}
{2 [ - (\Omega-\Omega_{\rm p})^2 +  B  \Omega  ]}.
\label{utheta}
\end{equation}

\subsection{Dissipation}

We now consider the additional dissipation in the disc due to tidal
perturbations when we introduce a small amount of dissipation into the
flow. The two modes of dissipation in a Newtonian fluid are the shear
viscosity, $\nu$, and the bulk viscosity, $\zeta$.  The dissipation
due to the bulk viscosity per unit radius is
\begin{equation}
D_1=\Sigma \zeta \int_0^{2\pi}\left( \bm{\nabla \cdot u}\right)^2\, d\theta
\end{equation}
which we can evaluate using
equations (\ref{ur}) and (\ref{utheta}).  The dissipation due to the shear
viscosity is
\begin{equation}
D_2=2\Sigma \nu \int_0^{2 \pi} E_{ij}E_{ij} \, d\theta,
\end{equation}
where $E_{ij}=e_{ij}-1/3\delta_{ij} \bm{\nabla.u}$ and $e_{ij}$ is the
rate of strain tensor which is given by
\begin{align}
e =\left(\begin{array}{c c}
\frac{\partial u_r}{\partial r}  & \frac{1}{2}\left[ r \frac{\partial (u_\theta/r)}{\partial r}+\frac{1}{r}\frac{\partial u_\theta}{\partial \theta}\right] \\
 \frac{1}{2}\left[ r \frac{\partial (u_\theta/r)}{\partial r}+\frac{1}{r}\frac{\partial u_\theta}{\partial \theta}\right]& \frac{1}{r}\frac{\partial u_\theta}{\partial \theta}+\frac{u_r}{r}\\
\end{array}\right).
\end{align}

\subsection{Torque on the disc}

From the dissipation we can find the torque on the disc due to the
tidal perturbation. The rate of working per unit surface area of the
disc by a torque, $T$, is
\begin{align}
D & =\Omega \frac{\partial T}{\partial r}dr \frac{1}{4\pi r dr} = \frac{\Omega}{4\pi r}\frac{\partial T}{\partial r} \cr
& =\frac{1}{4 \pi r}\left[\frac{\partial (T\Omega)}{\partial r}-T\Omega'\right],
\end{align}
where $\Omega'=d\Omega/dr$. The first term is the rate of convection
of energy over the whole disc. Its value depends only on the boundary
conditions. We find the torque on the disc to be
\begin{equation}
T_i=\frac{4\pi r D_i}{\Omega'},
\label{t}
\end{equation}
where $i=0$, $1$, $2$ \citep*{frank02}. The torque $T_0$ is defined as the torque on an
accretion disc without a companion. This has a viscous torque of
\begin{equation}
T_0=2\pi r \nu\Sigma r^2 \Omega'
\label{visct}
\end{equation}
and so the dissipation is
\begin{equation}
D_0=\frac{1}{2}r^2(\Omega')^2\nu\Sigma.
\end{equation}

\begin{figure*}
\centering
\includegraphics[width=8.4cm]{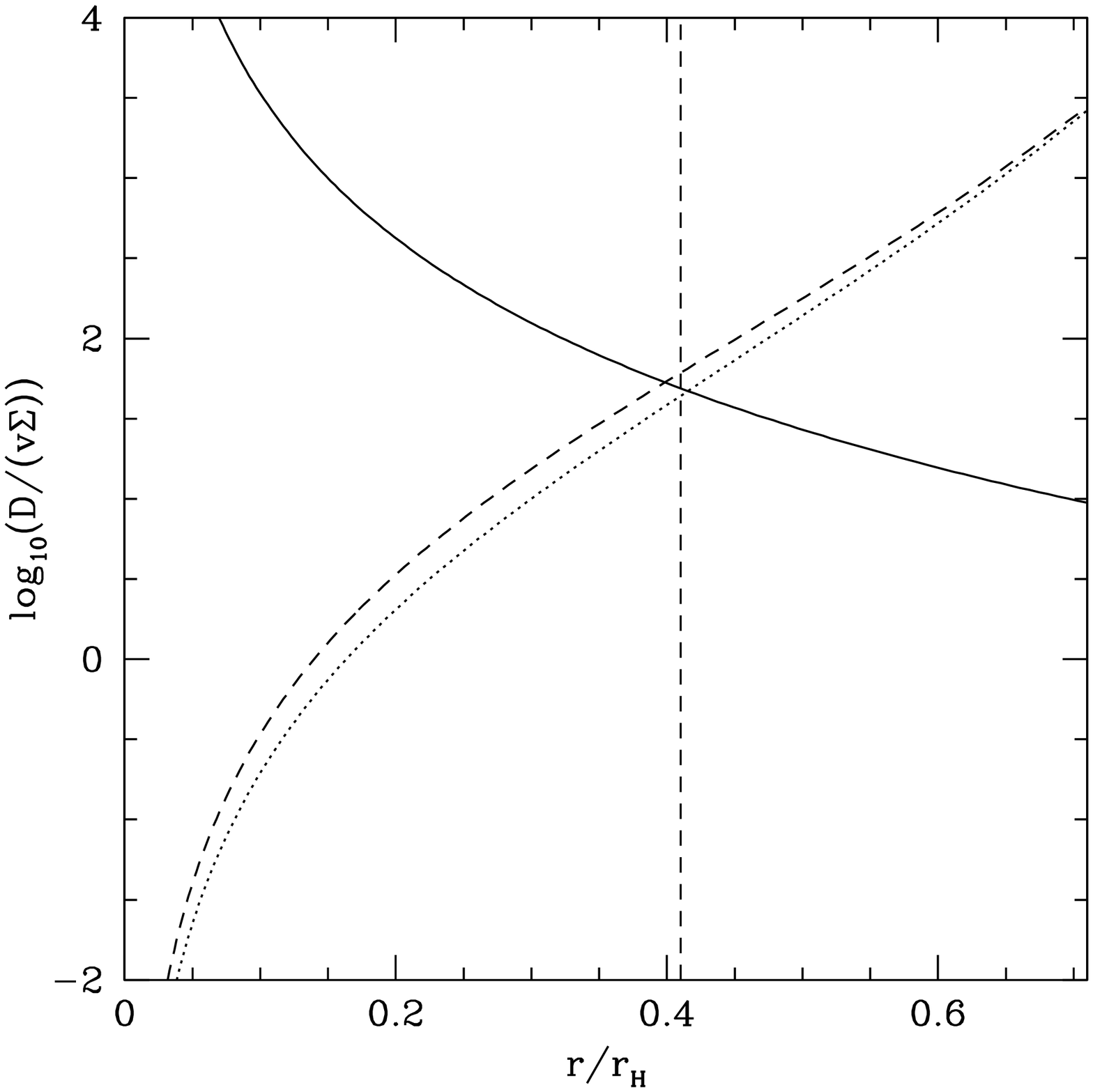}
\includegraphics[width=8.4cm]{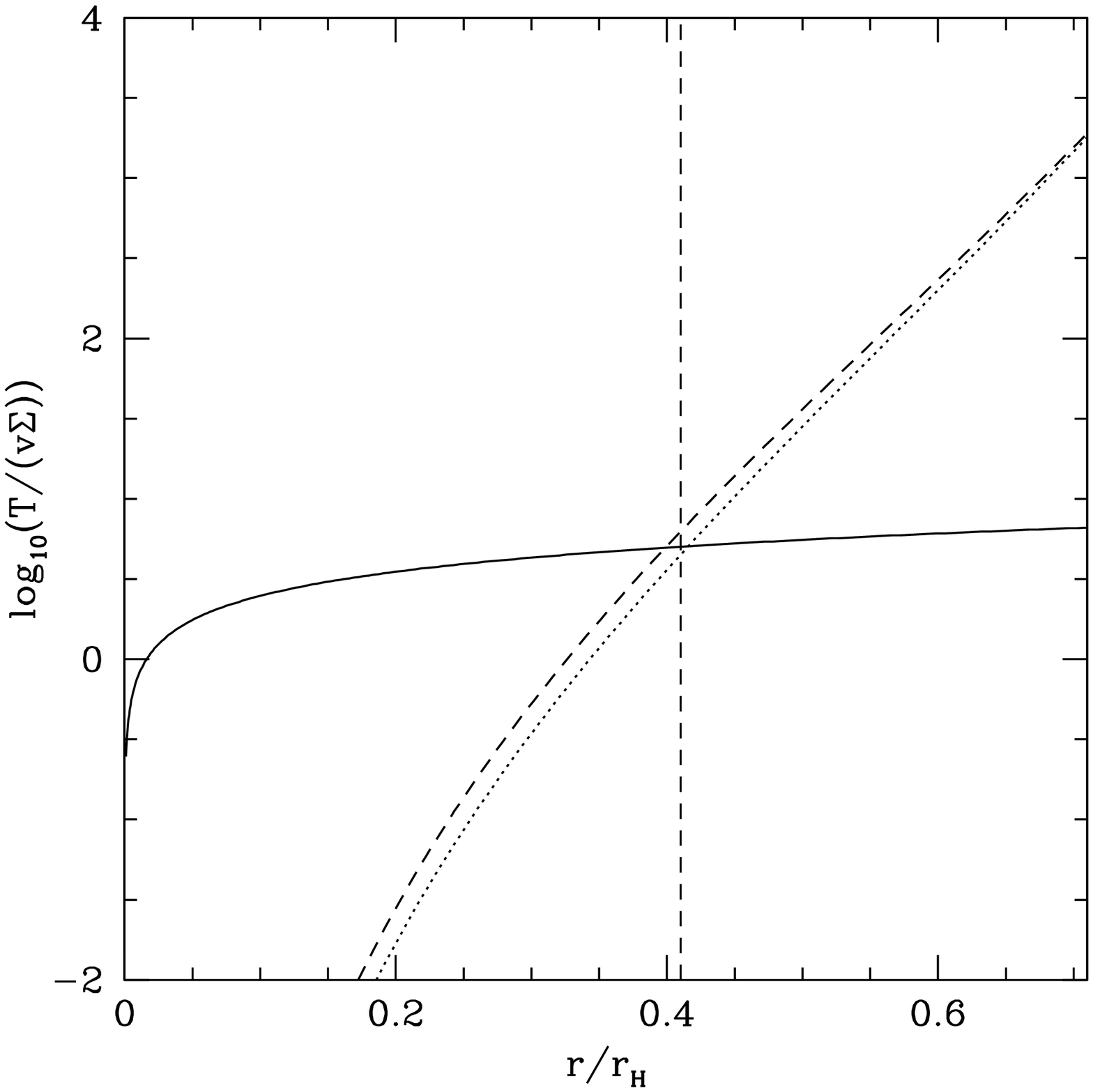}
\caption{Left: The dissipation in the disc. The solid line is
  $D_0/(\nu\Sigma)$, the dashed line is $D_1/(\zeta\Sigma)$ and the
  dotted line is $D_2/(\nu\Sigma)$. Right: The torque acting on the
  disc per unit surface density, per unit viscosity. The solid line is
  for the viscous torque in the disc, $T_0/(\nu\Sigma)$, the dotted
  line is the torque $T_1/(\zeta \Sigma)$ and the dashed line is
  $T_2/(\nu\Sigma)$. For both plots the vertical dotted lines show
  where the particle orbits cross.}
\label{diss}
\end{figure*}

In the left panel of Fig.~\ref{diss} we plot the three scaled
dissipations, $D_0/(\nu\Sigma)$, $D_1/(\zeta\Sigma)$ and
$D_2/(\nu\Sigma)$ as functions of the radius in the disc. This is similar
to Fig.~1 in \cite{papaloizou77}, but we use the Hill approximation.
We see that the magnitude of the dissipation from the internal viscous
torques, $D_0/(\nu\Sigma)$, and the tidal dissipation,
$D_2/(\nu\Sigma)$, are equal very close to, but just inside, the radius
in the disc where the particle orbits begin to cross. In the right
panel of Fig.~\ref{diss} we plot the scaled torques on the disc. The
tidal torque in the disc starts to dominate the viscous torque just
inside of the radius where the particle orbits cross.

However, outside of the radius where the particle orbits cross,
pressure and nonlinear effects cannot be neglected. Therefore,
this linear solution is not valid beyond the orbit crossing
radius. 

\subsection{One-Dimensional Simulations}
\label{oned}

We use a one-dimensional (radius only) model of an accretion disc
subject to a tidal torque in order to determine the surface density
evolution. The governing equation for the surface density of a flat
accretion disc centered on the planet is
\begin{equation}
\frac{\partial \Sigma}{\partial t}=\frac{1}{r}\frac{\partial}{\partial
  r} \left[ \frac{1}{[r^2 \Omega]'}
  \frac{\partial }{\partial r}\left(\nu \Sigma r^3 (-\Omega')\right)
 -\frac{\Sigma r  }{ [r^2 \Omega]'}  \frac{d T_{\rm gr}}{d M} \right] + S(r)
\label{sigmaev}
\end{equation}
\citep{pringle81}, where $'=d/d r$ and the angular velocity is given
in equation~(\ref{angvel}). In the previous section, we found the
tidal torque on the disc. However, this is only valid up to the
position in the disc where the orbits cross. We adopt a torque
function per unit disc mass on the disc of the form
\begin{equation}
\frac{ d T_{\rm{gr}}}{d M}= -T_3(r) \left(\frac{r}{r_{\rm H}}\right)^g,
\label{Tgr}
\end{equation}
where  $g$ is a constant.  To model the tidal effects of
orbit crossings, we select the torque parameters so that the torque
acts in the region where the particle orbits begin to cross.  We
choose $g=4$ so that the disc is truncated quickly.  The form of
$T_3$ controls where the  disc is truncated. We choose
\begin{equation}
T_3(r)=\begin{cases}0.5 \, r_{\rm H}^2 \,\Omega_{\rm p}^2 & r \ge 0.4\,r_{\rm H}\cr
0 & r<0.4\,r_{\rm H}.\end{cases}
\label{T0}
\end{equation}
To model the effects of the inflowing circumstellar gas, we model gas injection
at some radius $r_{\rm inj}$  over a narrow region of radial width $2 w$ 
at a steady rate $\dot M_{\rm inj}$ with the local Keplerian
speed. Function $S(r)$ describes the mass injection that we take to
be
\begin{equation}
S(r) = \frac{\dot M_{\rm inj}}{2 \pi r_{\rm inj}} \frac{H((r-r_{\rm inj})/w)}{2 w},
\end{equation}
where $H(x)$ is unity for $|x| < 1$ and zero otherwise.  We adopt a
width $w = 0.0046 \,(r_{\rm inj}/r_{\rm H})^\frac{1}{2}\, r_{\rm H}$.

We solve this equation numerically on a fixed mesh that is uniform  $r^\frac{1}{2}$ with 200 grid points
\citep[like][]{martin07}. We choose zero torque boundary conditions at
the inner boundary $r_{\rm in}=10^{-3}\,r_{\rm H}$ and the outer
boundary $r_{\rm out}=0.9\,r_{\rm H}$. The inner boundary allows the
material there to be accreted by the planet. Its position has been
chosen so that it is about equal to the radius of a Jupiter planet at
a distance of $5\,\rm AU$ from a central solar mass star. With the
tidal torque acting, the outer boundary is far enough out that it does
not affect the evolution because the tidal torque prevents the mass
from reaching the outer regions. However, if there is no tidal torque,
the disc extends as far as it can and the outer boundary will affect
the mass in the disc because mass is removed there. The viscous torque
is given in equation~(\ref{visct}) and we take $\Sigma=0$ at the
boundaries to have zero torques there. We initially take the surface
density to be a constant but very small value and allow it to build up
by mass accretion at a radius $r_{\rm inj}$.

\begin{figure*}
\begin{center}
\includegraphics[width=8cm]{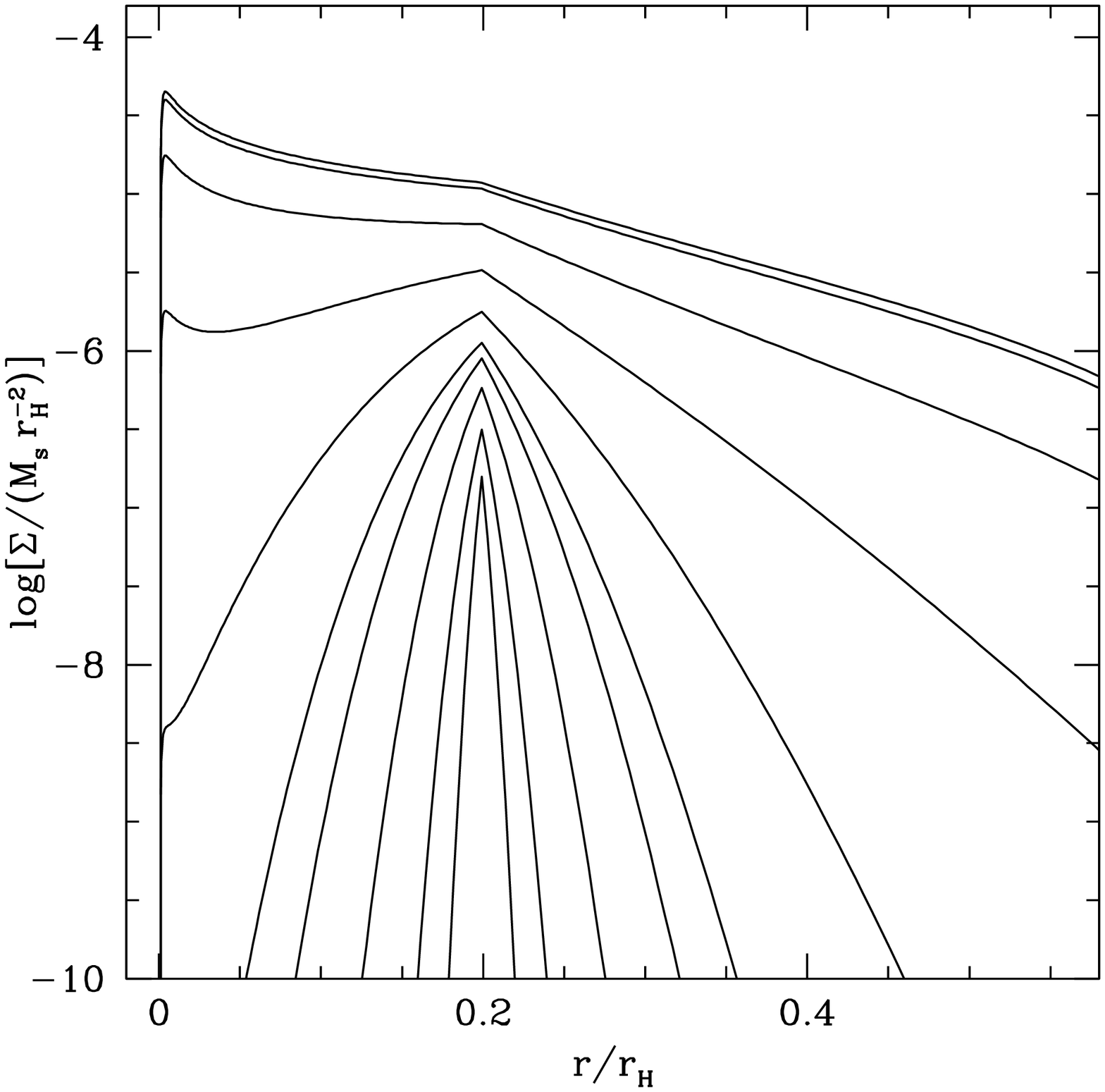}
\includegraphics[width=8cm]{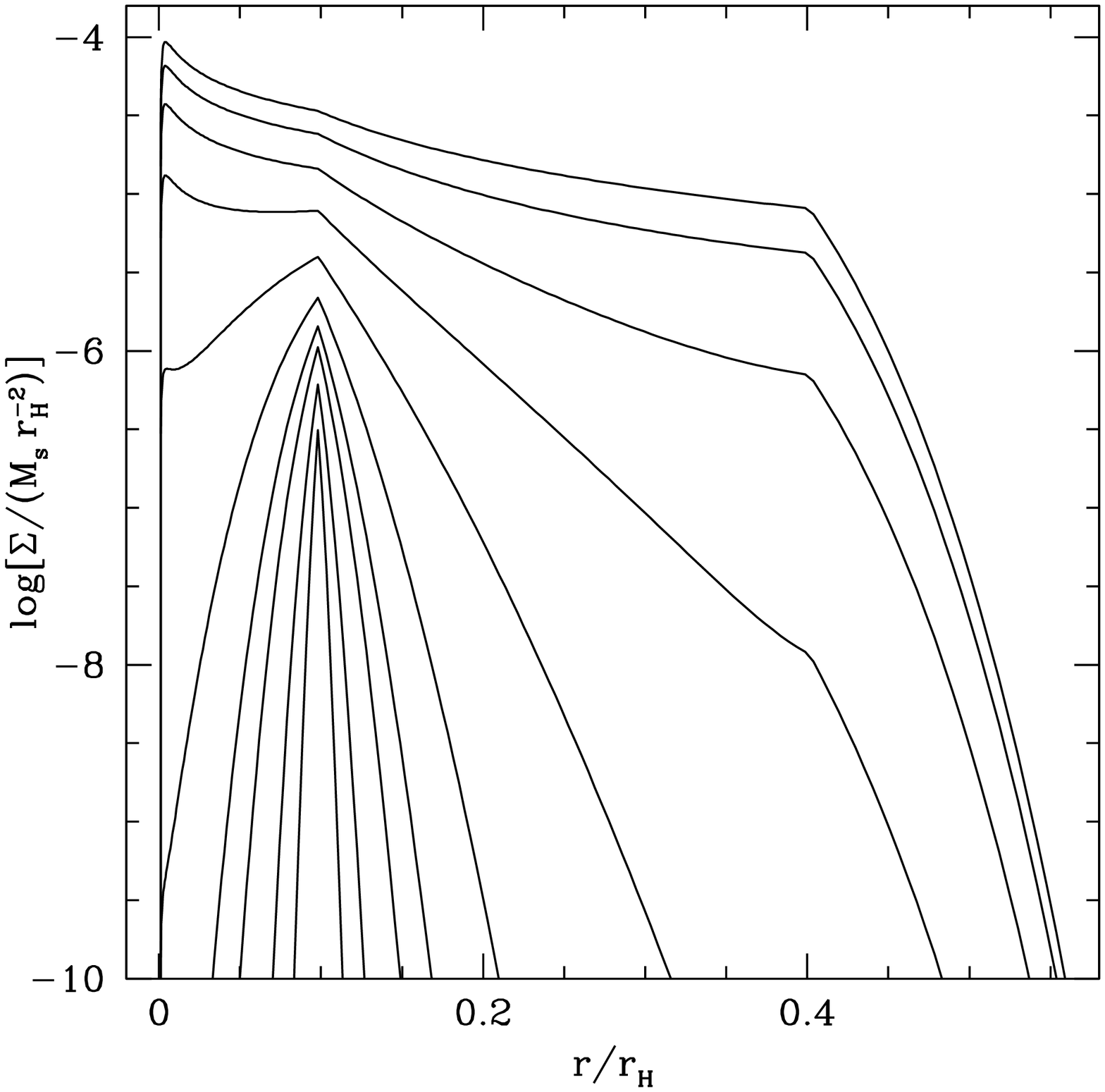}
\includegraphics[width=8cm]{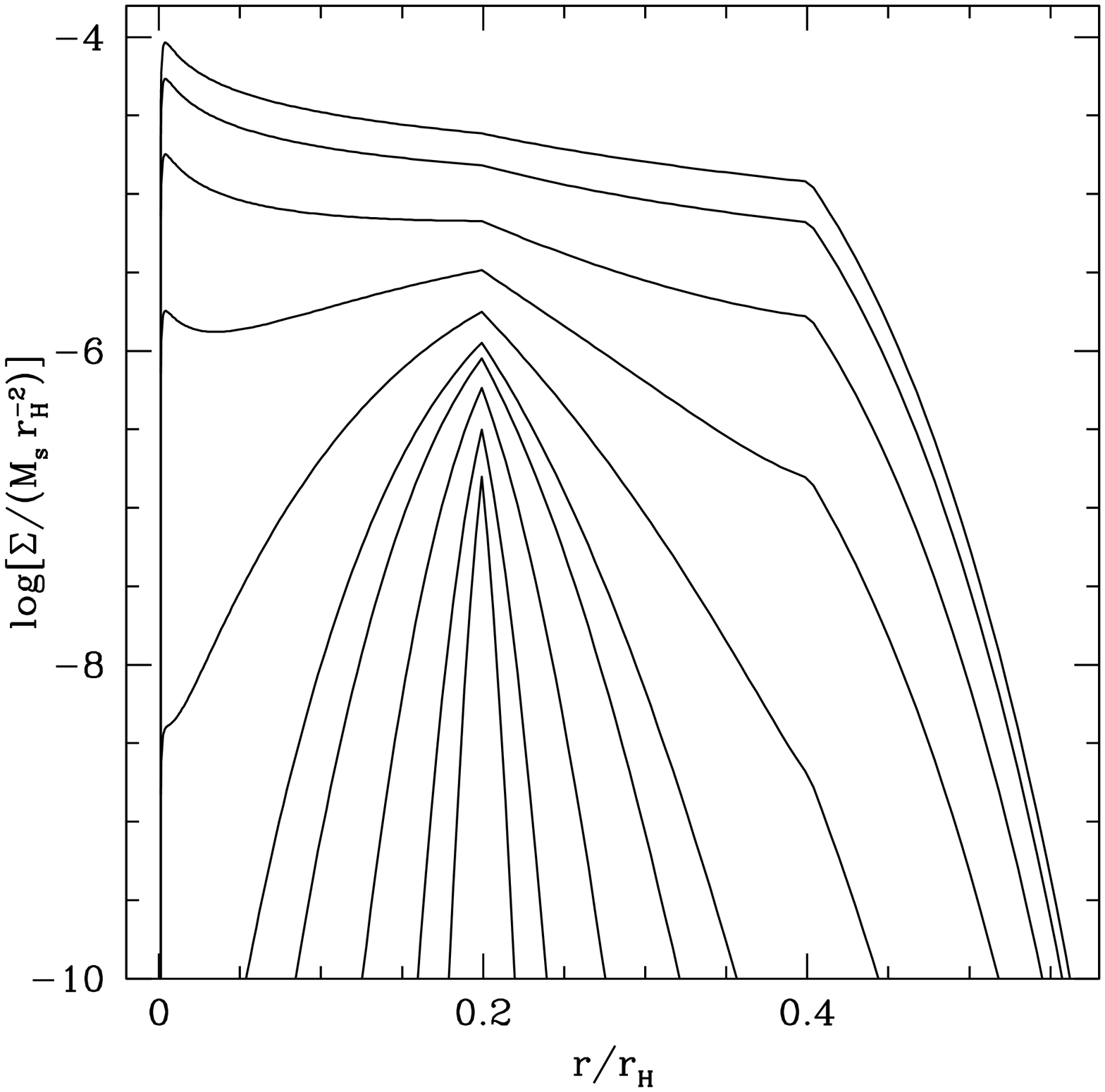}
\includegraphics[width=8cm]{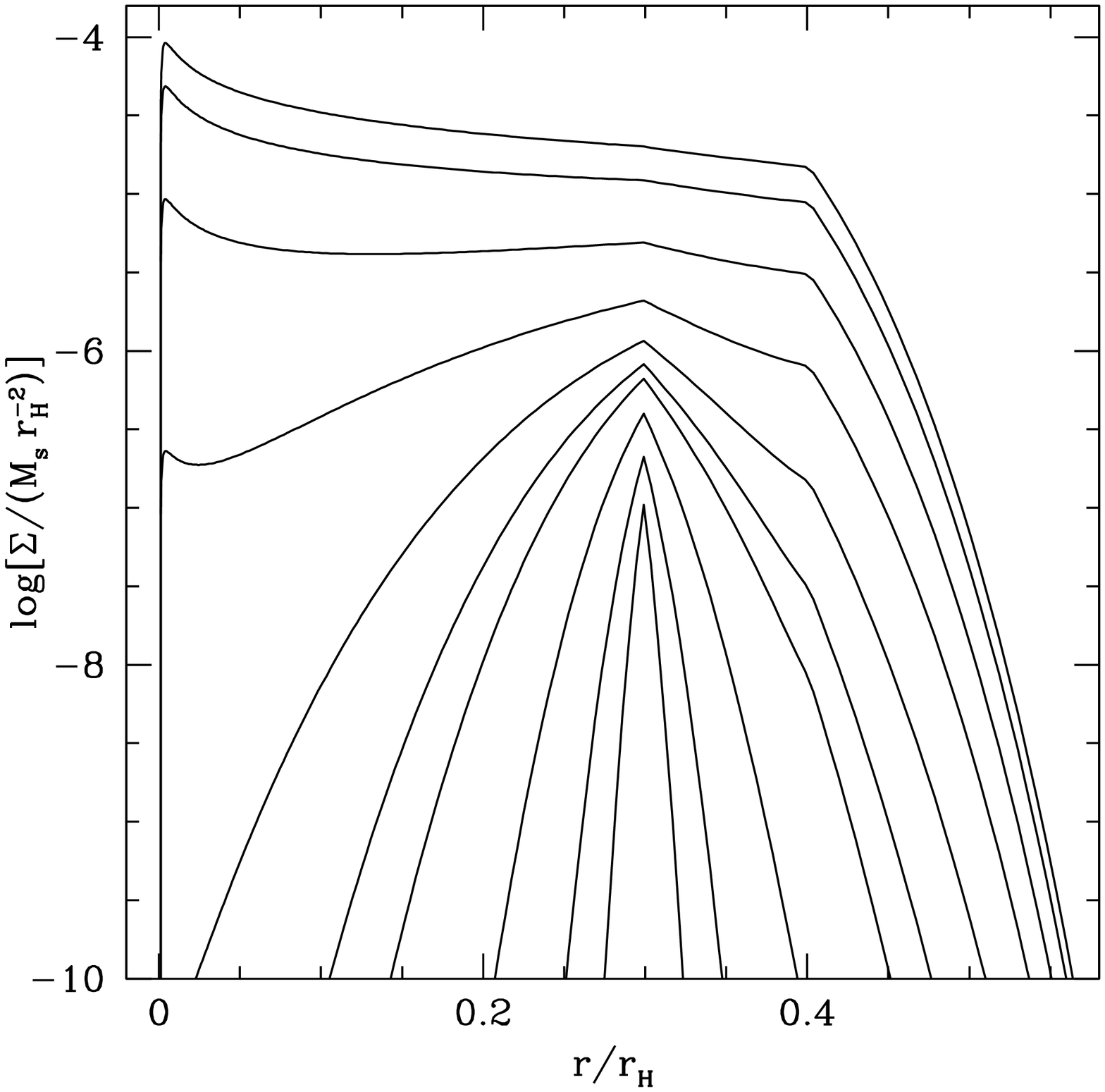}
\end{center}
\caption{Top left: The evolution of the disc without the tidal torque
  with $r_{\rm inj}=0.2\,r_{\rm H}$. The remaining plots show the
  evolution of the disc with the binary torque with $r_{\rm
    inj}=0.1\,r_{\rm H}$ (top right), $r_{\rm inj}=0.2\,r_{\rm H}$
  (bottom left) and $r_{\rm inj}=0.3\,r_{\rm H}$ (bottom right). The
  surface density increases in time. The times plotted, in order of
  increasing surface density in the plots, $t=0.01$, $0.05$, $0.18$,
  $0.42$, $0.67$, $1.63$, $5.50$, $20.96$, $82.79$ and $330.13\, P$
  where $P$ is the orbital period of the planet, $P=2 \pi/\Omega_{\rm
    p}$. }
\label{evolution}
\end{figure*}

We set the accretion rate on to the disc to be $\dot M=1.78 \times
10^{-8}\, M_{\rm s}\, \Omega_{\rm p}$. For a Jupiter mass planet
orbiting a solar mass star at a radius of $5\,\rm AU$ this corresponds
an accretion rate of $\dot M=10^{-8}\,\rm M_\odot\,yr^{-1} $. We start
with a disc of nearly zero mass. We note that changing the accretion
rate does not change the results in Fig.~\ref{evolution}, only the amount of mass in the disc.
We parametrise the viscosity with the $\alpha$-prescription in
equation~(\ref{alphavisc}). For a Keplerian disc with $\alpha=10^{-3}$
and a constant disc aspect ratio $H/r=0.3$, we have gas kinematic
turbulent viscosity
\begin{equation}
\nu=1.56\times 10^{-4} \left(\frac{r}{r_{\rm H}}\right)^\frac{1}{2}
r_{\rm H}^2\,\Omega_{\rm p}.
\label{nuv}
\end{equation}

Disc mass builds up as mass is injected.  We ran the numerical code
until it reached a steady state.  In Fig.~\ref{evolution} we plot the
surface density evolution of the disc for different injection
radii. The surface density in each of the plots increases as time goes
on.  The top left plot shows the evolution without the tidal torque
and an injection radius $r_{\rm inj}=0.2\,r_{\rm H}$. We see that the
disc spreads out as far as it can out to the outer boundary, where mass
is removed. All the other three plots in Fig.~\ref{evolution} include
the tidal torque and we vary the position that mass is added from
$r_{\rm inj}=0.1\,r_{\rm H}$ (top right), $0.2\,r_{\rm H}$ (bottom
left) and $0.3\,r_{\rm H}$ (bottom right).  With the tidal torque
acting, the disc cannot expand out to the outer boundary, it becomes
truncated well inside that boundary. The position of the injection of
the mass does not affect the outer boundary of the disc, it only
mildly modifies the surface density profile of the steady state
disc.

\subsection{One-Dimensional Analytic Solutions}
\label{analytic}

\begin{figure}
\begin{center}
\includegraphics[width=8cm]{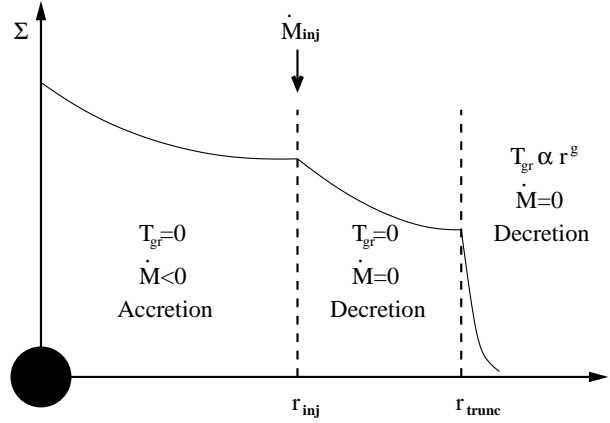}
\end{center}
\caption{A sketch of the surface density about a planet in the three region
  model. In $r<r_{\rm inj}$ there is an accretion disc and in
  $r>r_{\rm inj}$ there is a decretion disc structure that acts as a
  mass reservoir. In this model, for $r<r_{\rm trunc}$ there is no
  imposed gravitational torque and for $r>r_{\rm trunc}$ a power law
  gravitational torque is applied.}
\label{sketch}
\end{figure}

In this section we find steady-state analytical solutions for the
surface density of a circumplanetary accretion disc.  Our approach
is similar to that of  \cite{canup02} and  \cite{ward10}, but we include
the strong tidal torques near the orbit crossing radius.
There are three
regions in the disc, as shown in Fig.~\ref{sketch}. Inside of the
radius where mass is added, $r_{\rm inj}$, the disc acts as a normal
accretion disc. Outside of this radius, the disc acts as a decretion
disc in terms of the density profile, although gas does not actually
flow outward, as shown below. Instead, this region acts as mass
reservoir.  However, material in this region is subject to exchange
with the interior region  $r< r_{\rm inj}$ by means of
turbulent diffusion. A power law torque is applied only in the outer parts of
the disc in $r>r_{\rm trunc}$. We assume for simplicity that the disc
is in exact Keplerian rotation about the planet with angular speed
$\Omega$ in the inertial frame.

The accretion rate through the disc is
\begin{equation}
\dot M=2\pi r u_r \Sigma.
\end{equation}
Notice that $\dot M$ is negative for accretion ($u_r <0$).
Mass is injected at a radius $r=r_{\rm inj}$ with rate $\dot M_{\rm inj}>0$, that is
\begin{equation}
\frac{d\dot M}{dr}=\dot M_{\rm inj} \delta(r-r_{\rm inj})
\label{mass}
\end{equation}
and the angular momentum equation is
\begin{equation}
 \frac{d( \dot M r^2 \Omega)}{dr}=\frac{d}{dr}(T_\nu+T_{\rm gr}) +\dot
M_{\rm inj} r_{\rm inj}^2 \Omega_{\rm inj} \delta(r-r_{\rm inj}),
\label{angularmom}
\end{equation}
where the viscous torque per unit radius (with equation~\ref{visct})
is given by
\begin{equation}
T_\nu=-3 \pi r^2   \Omega  \Sigma \nu
\end{equation}
for a Keplerian angular velocity where $\Omega_{\rm inj}=\Omega(r_{\rm
  inj})$ and $T_{\rm gr}$ is the gravitational torque. Equation
(\ref{angularmom}) is equivalent to equation (\ref{sigmaev}) with the
time derivative set to zero and injection width $w$ set to (nearly)
zero.

 In accordance with equation (\ref{Tgr}), the gravitational torque per
 unit radius on the disc is taken to be
\begin{equation}
\frac{dT_{\rm gr}}{dr}=-2 \pi r \, \Sigma \, T_3(r) \, \left(\frac{r}{r_{\rm H}}\right)^g.
\end{equation}
where $T_3$ is defined by equation  (\ref{T0}). Integrating equation~(\ref{mass}) we find
\begin{align}
\dot M = \begin{cases} 
-\dot M_{\rm inj} +C_1 & r<r_{\rm inj} \cr
C_1 & r>r_{\rm inj} ,
\end{cases}
\end{align}
where $C_1$ is a constant. With the boundary condition that $\dot M=0$
at the disc outer edge, we see that $C_1=0$. So there is no accretion
flow outside of the injection radius, all of the material that is
injected at $r_{\rm inj}$ flows inwards to be accreted on to the
planet.  We now integrate equation~(\ref{angularmom}) to find
\begin{align}
 T_\nu + \dot M_{\rm inj} r^2 \Omega  +C_2 =0 ,& \,\,\,\,\,\,\,\, r<r_{\rm inj}  \cr 
T_\nu +\dot M_{\rm inj} r_{\rm inj}^2
\Omega_{\rm inj}+C_2 =0, &  \,\,\,\,\,\,\,\, r_{\rm inj}<r<r_{\rm trunc} \cr 
T_\nu+T_{\rm gr} +\dot M_{\rm inj} r_{\rm inj}^2 \Omega_{\rm inj}+C_2 =0, &  \,\,\,\,\,\,\,\, r>r_{\rm
  trunc},
\label{acc}
\end{align}
where $C_2$ is a constant to be determined by the inner boundary
condition.  As in Section \ref{oned}, we have chosen the gravitational torque to be zero in $r<r_{\rm
  trunc}$.

In evaluating equation (\ref{acc}), we apply the form of kinematic
viscosity given by equation (\ref{nuv}), $\nu(r) \propto \sqrt{r}$,
that holds for a constant $\alpha$ and constant $H/r$ circumplanetary
disc.   With the zero viscous torque inner boundary
condition at an inner radius $r_{\rm in}$, equation~(\ref{acc}) requires $C_2 = -\dot M_{\rm inj}
r_{\rm in}^2 \Omega_{\rm in}$.  We then obtain the standard viscous
disc surface density in the inner parts of the disc $r<r_{\rm inj}$,
designated by subscript 1,
\begin{equation}
\Sigma_1(r) =\frac{\dot M_{\rm inj}}{3 \pi\nu(r)}\left[1-\left(\frac{r_{\rm in}}{r}\right)^\frac{1}{2}\right]
\end{equation}
\citep{pringle81}. Therefore, $\Sigma_1(r) \propto 1/\sqrt{r}$, for $
r_{\rm inj} > r \gg r_{\rm in}$.  In the other two regions of space
where $r> r_{\rm inj}$, we assume that $r \gg r_{\rm in}$, so that
$C_2$ is very small compared to the other torques. For simplicity, we
take $C_2=0$ in these regions.

In the region outside of where the mass is injected, but where no
gravitational torque is applied, $r_{\rm inj}<r<r_{\rm trunc}$, the
surface density is
\begin{equation}
\Sigma_2(r)= \Sigma_1(r_{\rm inj})\left(\frac{r_{\rm inj}}{r}\right).
\end{equation}
In the outermost region, $r \ge r_{\rm trunc}$, the surface density is
\begin{equation}
\Sigma_3= \Sigma_2(r_{\rm trunc}) \left(\frac{r_{\rm trunc}}{r}\right) \exp \left[ -\frac{2(r^{1+g}-r^{1+g}_{\rm
    trunc})T_3(r_{\rm trunc})}{3(1+g) \nu_{\rm trunc} \Omega_{\rm trunc} r_{\rm
  trunc} r_{\rm H}^{g}}\right],
  \label{sig3}
\end{equation}
where $\nu_{\rm trunc}=\nu(r_{\rm trunc})$ and $\Omega_{\rm
  trunc}=\Omega(r_{\rm trunc})$.  We now have the surface density for
the whole disc with the equations for $\Sigma_1$, $\Sigma_2$ and
$\Sigma_3$.  We compared these solutions to the one-dimensional cases
of Section \ref{oned} with the same values for the input parameters.
These steady state analytic solutions agree well (better than 6\%)
with the numerical solutions obtained at late times, plotted as the
highest lines in the plots in Fig.~\ref{evolution}.

\begin{figure*} 
\centering
\includegraphics[width=5.0cm]{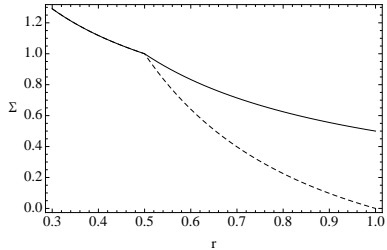}
\caption{Surface density distributions for a disc with constant $\alpha$ and constant $H/r$ as a function of radius for cases
where the mass is injected at radius $r=0.5$. 
The surface density is normalized
by its value at the injection radius. The dashed line is for a model that satisfies outer boundary
condition $T_\nu =0$ at the disc outer radius $r_d$. The unit of radius on the plot is $r_d$ in the case.
The solid line is for a case that models the effects of the tidal field. The tidal field acts to sharply
truncate the disc outside of the truncation radius $r_{\rm trunc}$. The density distribution in this case satisfies
satisfies boundary condition (\ref{tobc}) and the unit of radius in the plot is $r_{\rm trunc}$.
The region $r>  r_{\rm trunc}$, where the density drops to zero, is not shown.
There are additional corrections to this profile due to the effects of gas pressure in the outer
parts of the disc. }
\label{dcomp}
\end{figure*}

\cite{canup02} have also considered the flow outside some radius where
mass is injected, but did not include the effects of the tidal
field. They have a normal accretion disc in the inner parts, as we
have. But in the outer parts of the disc in their model, the material
flows outwards and is removed from the disc at a rate $0.8\, \dot
M_{\rm inj}\sqrt{r_{\rm inj}/r_{\rm d}}$ where $r_{\rm d}$ is the
outer edge of disc. For their standard model with $r_{\rm
  inj}=30\,r_{\rm J}$ and $r_{\rm d}=150\,r_{\rm J}$  (for Jupiter radius $r_{\rm J}$),
  we find the mass
loss rate is $0.36\,\dot M_{\rm inj}$.  If we take the disc radius to be the orbit crossing
radius, the mass loss rate is about $0.25\,\dot M_{\rm inj}$.
 They suggest
that the removal of matter at the outer edge of the disc occurs
through solar torques or collisions with highly shocked
regions. However, in our model the tidal torque dominates the viscous
torques in the outermost regions and prevents the outward flow of
matter. Hence, in our picture of a steady state, the rate that mass is
accreted on to the central planet is the same rate that it is injected
into the disc. In addition, there is a  difference in the
 density profiles between the models outside the mass injection radius.
 In that region,  \cite{canup02} 
determined
 a gradual tapering that varies in radius as $\sqrt{r_{\rm d}/r}-1$ for disk radius $r_{\rm d}$. 
  In our model, the density
falls off more slowly in radius in the main body of the disc and is abruptly
terminated near the orbit crossing radius. The difference in
the density structures can be traced to outer boundary conditions.
\cite{canup02}  employed a zero viscous stress outer boundary condition, $T_\nu=0$. 
In our model, the effects of the tidal field can be approximately represented
by an outer boundary condition at the tidal truncation radius (orbit crossing radius) that is given by
\begin{equation}
T_\nu = -\dot M_{\rm inj} r_{\rm inj}^2 \Omega_{\rm inj},
\label{tobc}
\end{equation}
 as follows from the second relation of equation (\ref{acc}) with $C_2=0$.
 Fig.~\ref{dcomp} compares the density distributions resulting from the two boundary
 conditions.
The density distribution we obtain approaches the
\cite{canup02} distribution in the limit that the injection radius goes to zero.
There are further modifications to the density profile due to pressure effects that
are described in Section~\ref{concs}.

From equation (\ref{sig3}), the width of the disc outer edge (ignoring
pressure effects) is estimated as
\begin{equation}
w = 1.5 \nu  \Omega  r \, \left | \frac{dT_{\rm gr}}{d M} \right |^{-1},
\label{w}
\end{equation}
where the right-hand side is evaluated near the outer edge.
If the disc flow is smooth and there are no resonances,
the phase lag that produces a tidal torque due to the turbulent
viscosity is estimated as $\delta \theta \sim \nu/(r^2 \Omega)$, the
inverse Reynolds number of the flow. The resulting torque
per unit disc mass is then estimated as $|dT_{\rm gr}/d M| \sim \delta \theta \, |\Phi_2|^2/(r^2 \Omega^2)$,
where $\Phi_2 $ is the $m=2$ component of the tidal potential.
Using equation (\ref{w}) and assuming $|\Phi_2| \ll r^2 \Omega^2$, it follows that $w \gg r$. Therefore, this torque
is not capable of  truncating the disc. Instead, the much stronger tidal
torque, due to orbit crossings and certain resonance effects discussed in Section \ref{concs},
 truncates the disc in a small region of space. Torques from the region well inside
of where orbit crossings occur are then too weak to truncate the disc. 

The analytic solutions can also be considered to apply when the matter
being injected into the disc has a nonKeplerian velocity, that is
$\Omega_{\rm inj}$ at $r_{\rm inj}$ is nonKeplerian.  The gas in such
a case adds angular momentum per unit mass to the system at a rate
that differs from the Keplerian rate. In reality, such gas would
undergo a strong Kelvin Helzholtz instability with the disc. Since the
gas being added to the disc has a much lower density than the disc, we
would expect this gas to be rapidly entrained by the disc that remains
in nearly Keplerian rotation. An example of such a situation occurs
when coplanar inflowing gas meets the circumplanetary disc at the disc
outer edge.  We expect the inflowing gas in this case to be
subKeplerian. In such a case, the decretion disc or mass reservoir
region, as depicted in Fig.~\ref{sketch} is very small. Instead,
nearly entire disc behaves an accretion disc. The subKeperian 
 injected gas can be considered to impart a negative torque on
an otherwise Keplerian disc, in addition to the tidal torque.  On the
other hand, if much of the inflowing gas flows over the disc before
becoming entrained by the disc, then the injection radius will be
smaller than the disc outer radius, as in the cases plotted in
Fig.~\ref{evolution}.

\begin{figure*} 
\centering
\includegraphics[width=5.0cm]{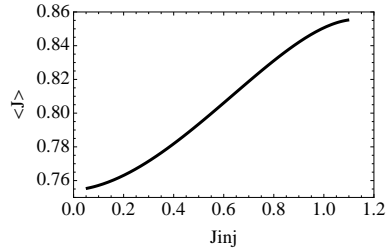}
\caption{Average specific angular momentum (angular momentum per unit mass) $\langle J \rangle$
 in a circumplanetary disc (total disc angular
momentum divided by disc mass), based on the steady state model of Section 6.5,  
plotted as a function of the specific angular momentum  
 of the accreting gas $J_{\rm inj}$. The quantities
on the horizontal and vertical axes are normalized by $r_{\rm H}^2 \Omega_{\rm p}$. 
The  range of the horizontal axis extends to  $J_{\rm inj}$ at the  disc outer edge, $r= 0.41 r_{\rm H}$.
The
results show that the average specific disc angular momentum is nearly independent of the 
specific angular
momentum of the accreting gas, even for small values. This model assumes that
the gas flows over the disc and is entrained vertically at the local Keplerian speed
for the mass injection radius. If planar subKeperian gas is added at the disc outer edge,
we expect the variations of $\langle J \rangle$ to be even smaller, since the gas distribution
within the main body of the disk behaves as a simple accretion disc.}
\label{jc}
\end{figure*}

Beyond the details of the flow properties in particular models, the
results suggest that the disc structure is generally largely
independent of the angular momentum per unit mass of the inflowing
gas, as seen in Figs.~\ref{evolution} and \ref{jc}.  The disc density distribution
is mildly influenced by this quantity. The radial derivative of the
surface density undergoes order unity changes at $r_{\rm inj}$. In the
model considered this section, the logarithmic surface density
gradient changes from -1/2 to -1 across $r_{\rm inj}$.  Also, for a fixed
mass injection rate, the density in the outer parts of the disc
scales as $\sqrt{r_{\rm inj}/r_{\rm trunc}}$. But such variations do
not lead to strong changes in disc structure by typically plausible values
of $r_{\rm inj} > 0.05 r_{\rm H}$.   
This is particularly
true if the injection radius is near the disc outer edge.  Instead,
the disc structure interior to the disc tidal truncation region is
dominated by the effects of disc turbulent viscosity.

\section{Disc Model With Pressure}
\label{sph}

\begin{figure*}
\centering
\includegraphics[width=8.4cm]{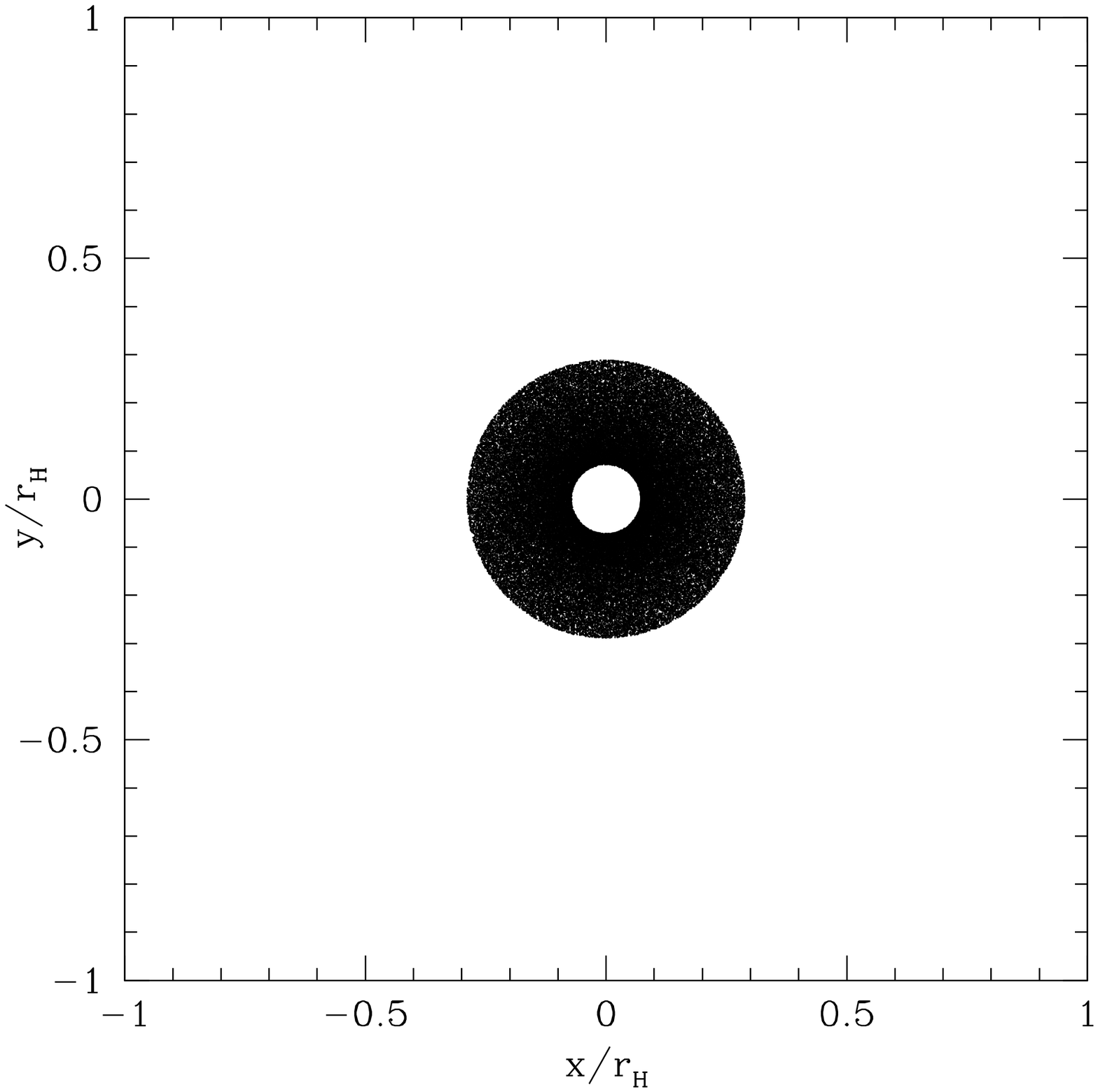}
\includegraphics[width=8.4cm]{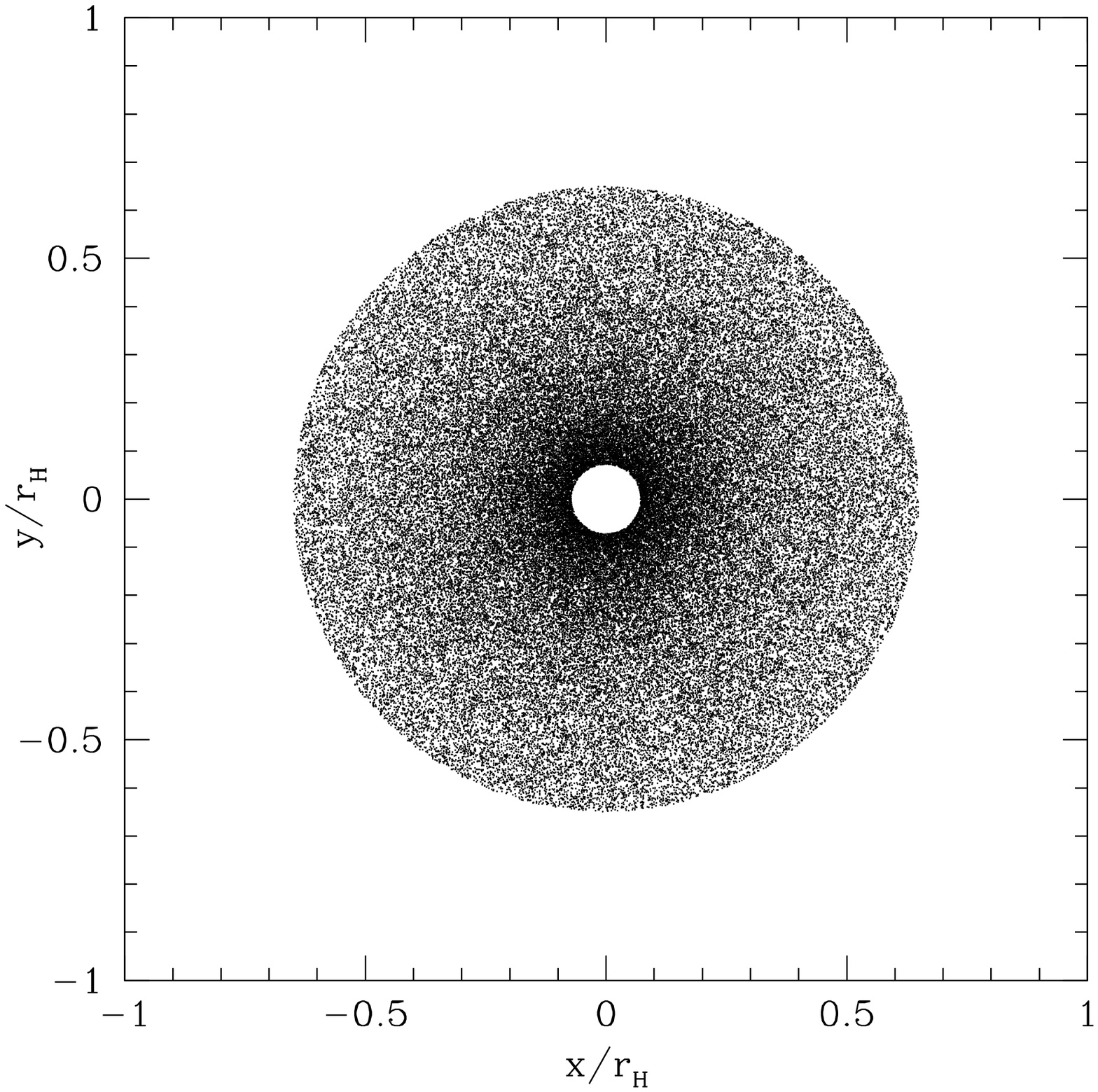}
\caption{ The initial configuration for the two SPH runs. Left: the
  particles are randomly distributed in radius of $0.07 r_{\rm H} < r <
  0.3 r_{\rm H} $. Right: The particles are randomly distributed
  randomly in angle and randomly in radius within and annulus defined
  by $0.07 r_{\rm H} <r < 0.65 r_{\rm H}$. }
\label{initial}
\end{figure*}

To include the effects of gas pressure, we modeled the disc using a
two-dimensional SPH code with $10^5$ particles.  The SPH parameter
$\alpha_{\rm SPH}$ was set to unity in these simulations, while
$\beta_{\rm SPH}$ was set to zero. The disc sound speed was crudely
modeled as a constant equal to $0.3\, \mu^{1/3} a \Omega_{\rm p}
=0.43\,\Omega_{\rm p} r_{\rm H} $ throughout. The disc aspect ratio
was $H/r \simeq 0.16$ at $R = 0.4 R_{\rm H}$ (in dimensionless units of these equations
$R_{\rm H} = 3^{-1/3}$), where free particle
orbit crossings occur.  The fluid equations were taken in the Hill
approximation, following the force equation (\ref{origs}).  The
particles were initially distributed so that surface density was
initially $\Sigma(r, \theta) \propto 1/R$ in an annulus $R_{\rm
  in}<R<R_{\rm out}$. We chose $R_{\rm in}=0.07 R_{\rm H}$ and have
two models, one with $R_{\rm out}= 0.3 R_{\rm H}$ and the second with
$R_{\rm out}= 0.6 R_{\rm H}$ which we plot in Fig.~\ref{initial}. The
initial velocities were taken to be circular rotation having angular
speed $\Omega - \Omega_{\rm p}$ in the corotating frame, with $\Omega$
given by equation (\ref{angvel}). Particles are removed from the
simulation if they reach the inner boundary at $R=0.07 R_{\rm H}$ or
the outer boundary located at $R=0.86 R_{\rm H}$. For each particle
removed, a particle is injected at a random angle and random radius
between $R=0.22 R_{\rm H}$ and $R=0.36 R_{\rm H}$.  Therefore, the
number of particles is fixed at $10^5$ at all times.

\begin{figure*}
\centering
\includegraphics[width=8.4cm]{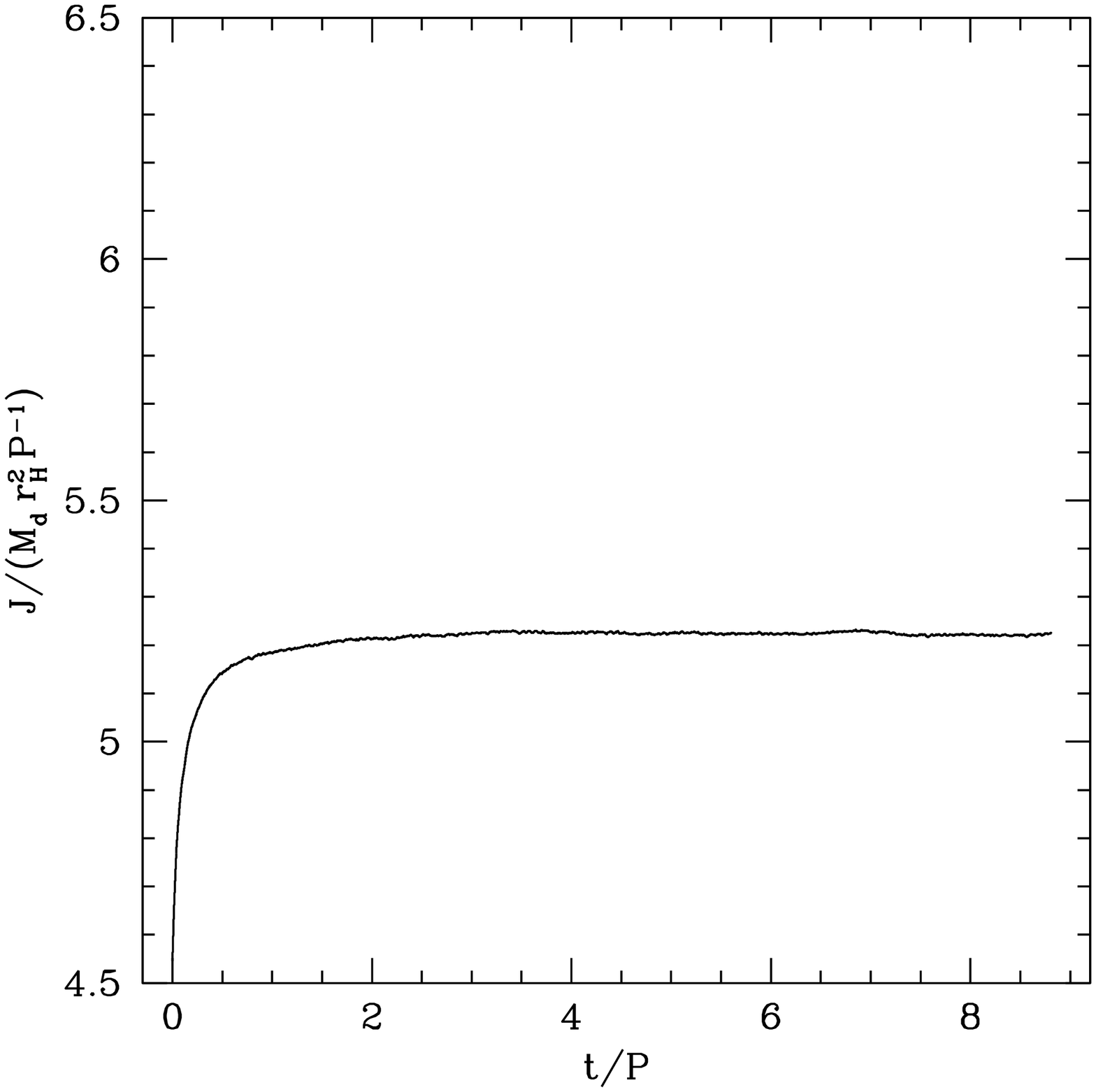}
\includegraphics[width=8.4cm]{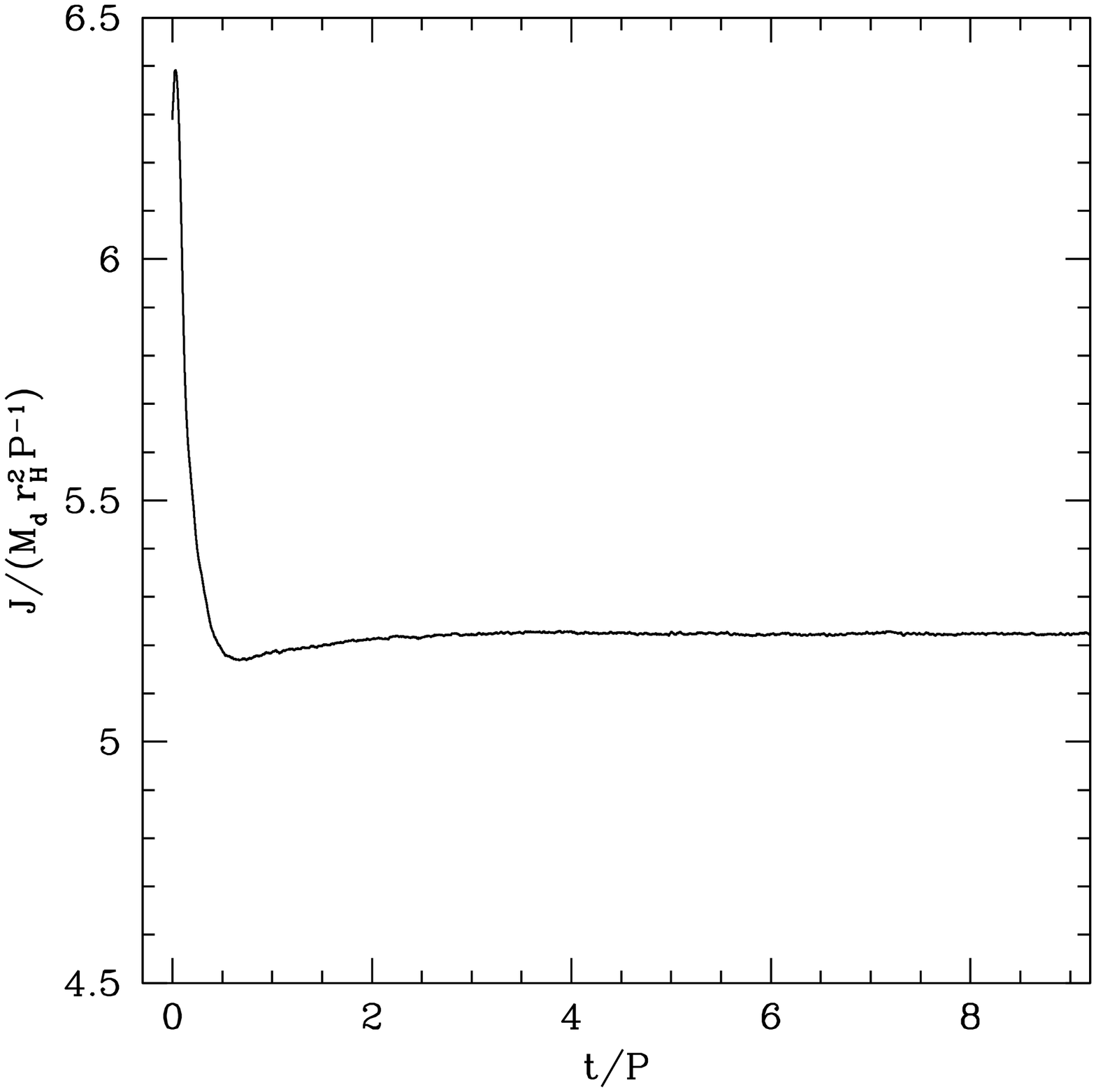}
\caption{The angular momentum evolution in time for the initially
  small disc (left) and the large disc (right), plotted in
  Fig.~\ref{initial}. The units of angular momentum are $ M_{\rm d}
  r_{\rm H}^2 P^{-1}$, for a disc of mass $M_{\rm d}$. The
  unit of time is the planet orbit period, $P$. }
\label{jevolution}

\end{figure*}

\begin{figure*}
\centering
\includegraphics[width=8.4cm]{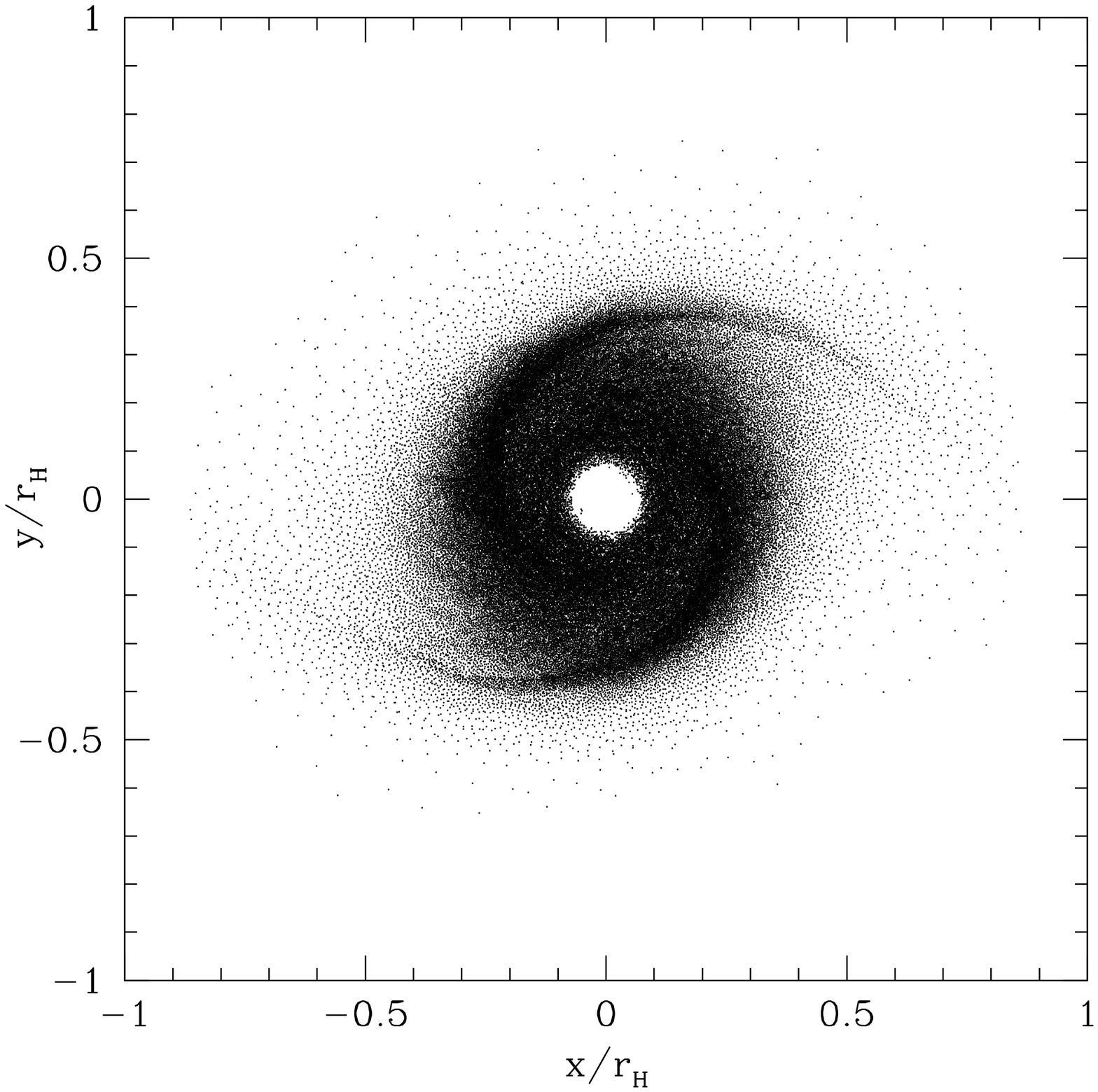}
\includegraphics[width=8.4cm]{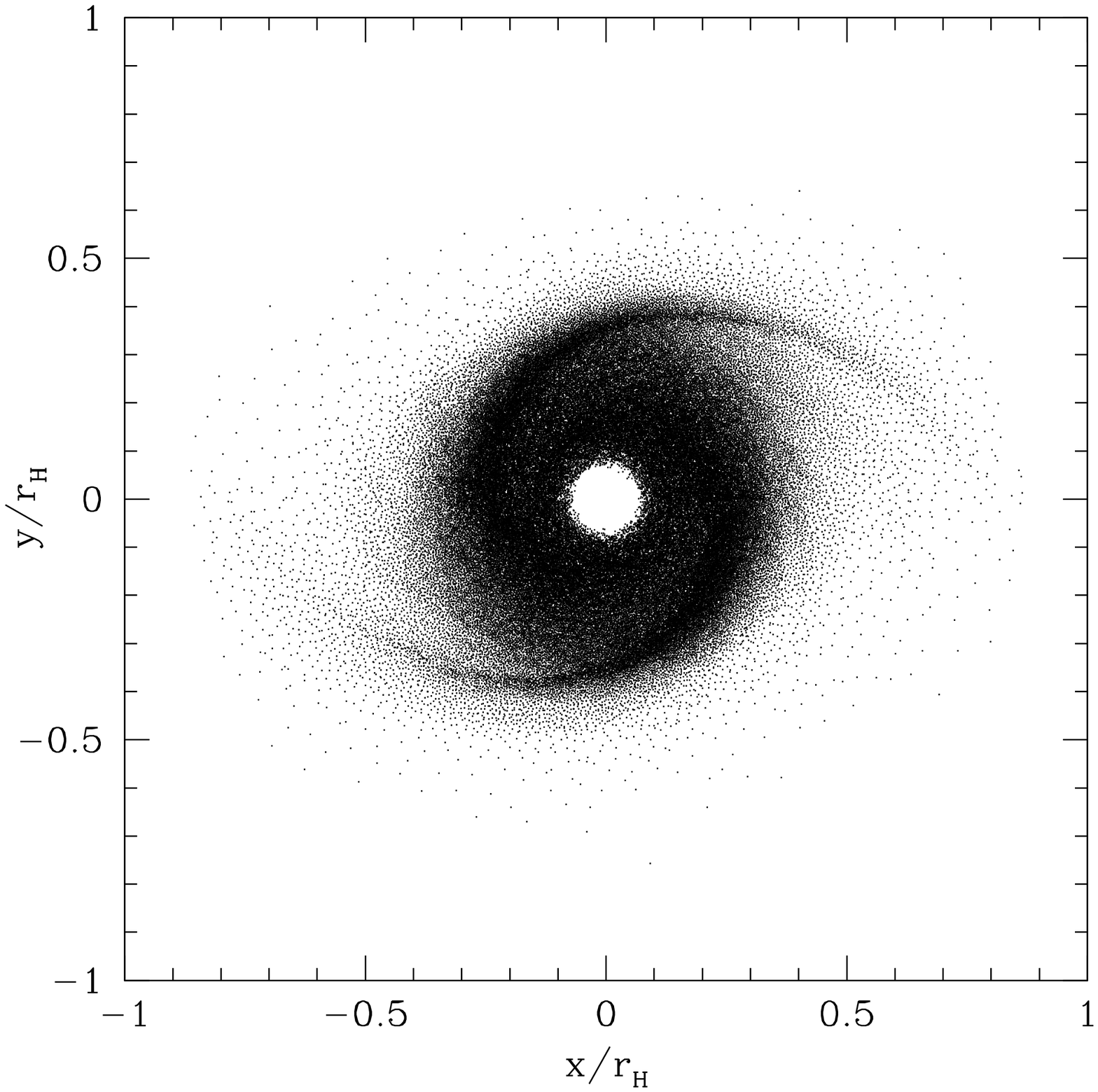}
\caption{ The particle distribution in space for the two SPH simulations
  corresponding to those with the initial configuration in
  Fig.~\ref{initial} at a time of 6 planetary orbital periods. }
\label{final}
\end{figure*}

Fig.~\ref{jevolution} plots the angular momentum evolution as a
function of time for the two initial disc sizes. The smaller disc
starts with much less angular momentum than the larger one. But after
only about 3 planetary orbits both discs have similar values of
angular momentum and approach a steady state.  In Fig.~\ref{final} we
plot the particles at a time of 6 planet orbital periods. We see that
particle distributions look very similar and the disc has reached a
near steady state. The viscous timescale of the disc can be estimated
as $\sim r^2/\nu$. In SPH, we have that $\alpha \simeq 0.1 \alpha_{\rm SPH} =0.1$
\citep[e.g.,][]{artymowicz94}. The value of the kinematic viscosity $\nu$ is
estimated as $0.1 \alpha_{\rm SPH} (H/r)^2 r^2 \Omega$ and
the viscous timescale evaluates to about 8 orbits at $R \simeq 0.3 R_{\rm H}$. 
The simulated disc has therefore settled to a near
equilibrium state on a timescale that is of order the estimated
viscous timescale.  This kinematic viscosity is a factor of ten or
more larger than what is typically taken taken in circumplanetary disc
simulations \citep[e.g.,][]{dangelo02, ayliffe09}. 
Consequently, in those simulations, the timescales to reach a steady
state are longer than the case here by similar factor. (In addition, previous simulations
have  generally not started with a circumplanetary disc. Some 
time is required for its formation from inflowing gas.) For the
simulations in \cite{machida09}, the timescale would be determined by
the inherent viscosity in the code due to the finite differencing,
since no viscous terms were included.

\begin{figure*} 
\centering
\includegraphics[width=7.0cm]{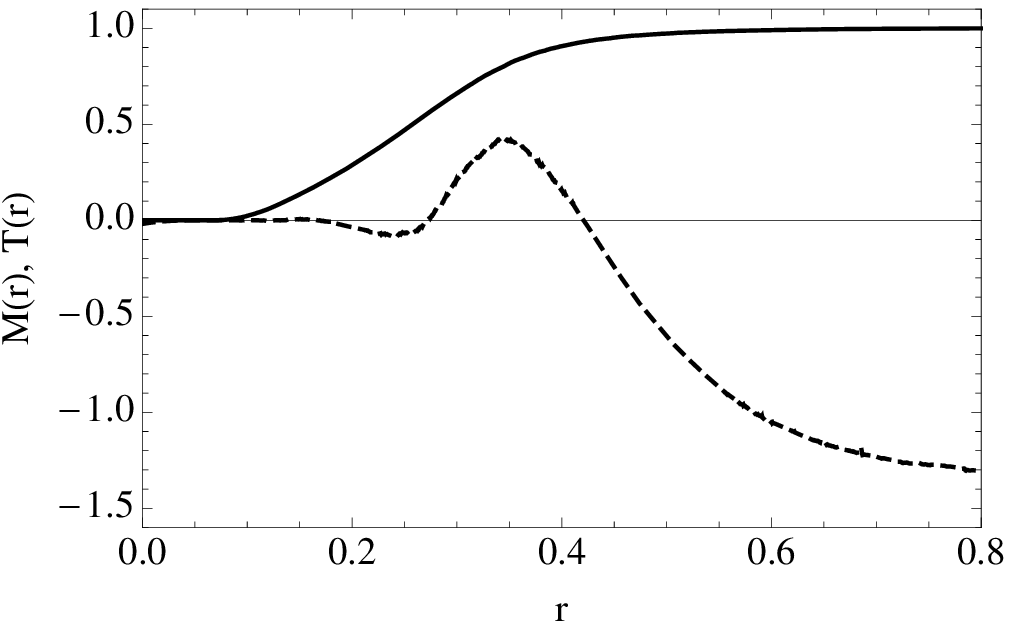}
\includegraphics[width=5.0cm]{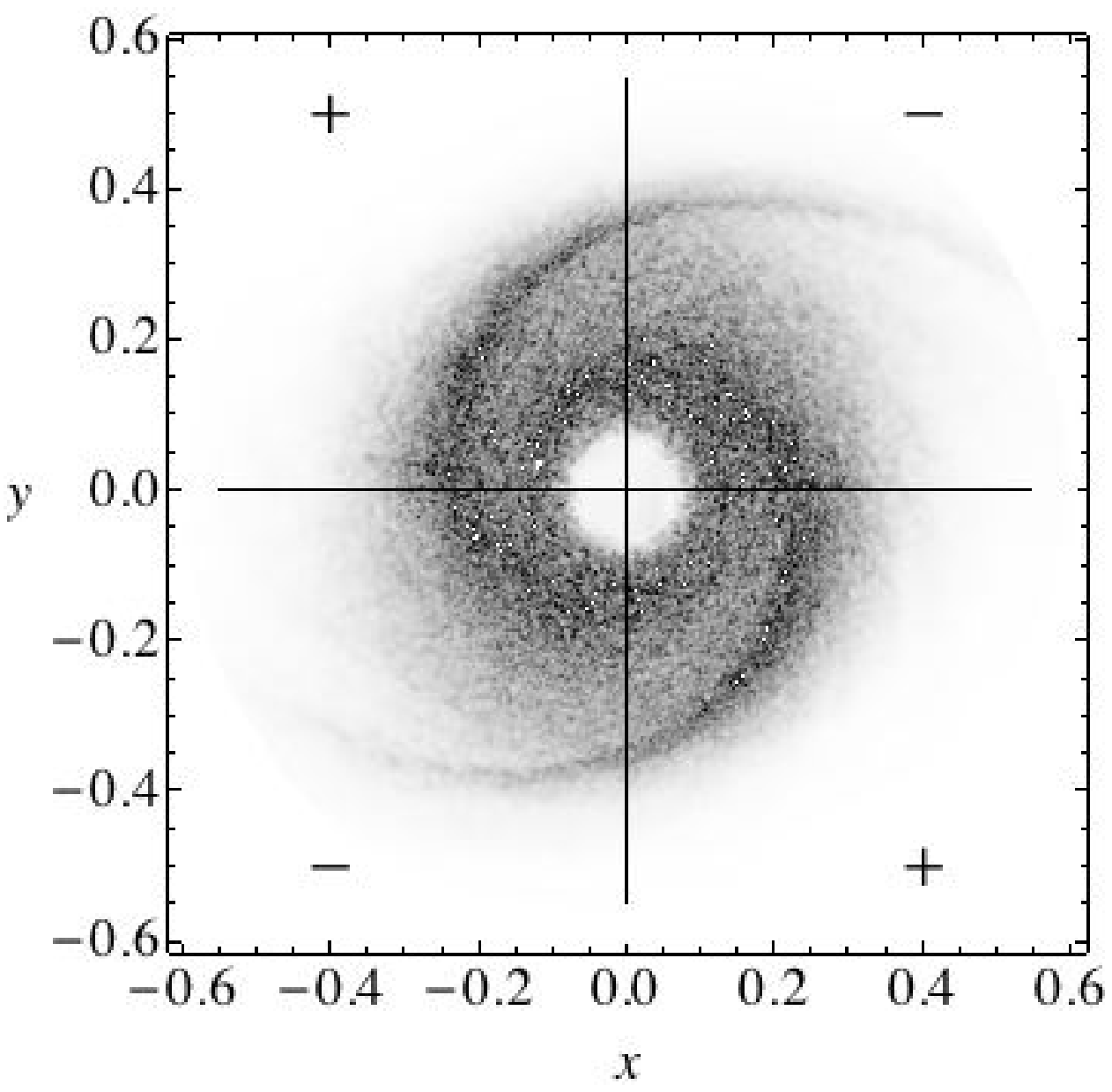}
\caption{Left: Cumulative disc mass (solid) in units of total disc
  mass $ M_{\rm d}$ and cumulative gravitational torque (dashed) on
  the disc in units of $0.01 M_{\rm d} r_{\rm H}^2 \Omega_{\rm p}^2$
  starting at the disc center as a function of radius in units of
  $r_{\rm H}$ at a time of 6 planetary orbits. Right: Plot of the
  particle positions in units of $r_{\rm H}$ with the point size
  weighted by density, in order to enhance the visibility of the
  spiral wave in the main body of the disc. The sign of the
  gravitational torque in each quadrant is also shown.  At small
  radii, $r \simeq 0.2$, the spiral lies in the first and third
  quadrants where the gravitational torque on the disc is negative,
  and the cumulative gravitational torque begins as negative in the
  left panel. At $r\simeq 0.3$, the spiral lies in the second and
  fourth quadrants where the gravitational torque on the disc is
  positive, and the cumulative gravitational torque turns
  positive. The outer parts of the spiral lie in the first and third
  quadrants and provide a negative gravitational torque contribution
  that causes the total torque on the disc to be
  negative.}
\label{phase}
\end{figure*}

The phasing of the arms determines the sign of the gravitational torque in the
different regions of space as shown in Fig.~\ref{phase}. The left plot
shows the cumulative gravitational torque as a function of radius in a
steady state disc after 6 planetary orbits.  On the right is a plot
of the shape of the spiral in the disc. The sign of the gravitational  torque is
equal to sign of $- x y$ and so is negative in the first and third
quadrants the torque and positive in the second and fourth.  As seen
in the left plot, the gravitational torque is negative in the outermost region of
the disc where it is strongest. The spiral there lies in the first and
third quadrants, as seen in Fig.~\ref{final}, in order the provide a net
negative gravitational torque that removes angular momentum from the disc as gas
accretes.

The left plot in Fig.~\ref{phase} also shows the cumulative disc mass
distribution.  More than 90\% of the disc mass is located inside a
radius of $0.41 r_{\rm H}$, where orbits crossings of free particles
occur. The local gravitational torque is negative for $r > 0.35 r_{\rm H}$.
On the other hand, the cumulative gravitational torque on the
disc inside the orbit crossing radius is actually positive. This does not mean that
the gas gains angular momentum in this region. The viscous torques
compensate by transporting the angular momentum out of this region so
that the overall torque is negative.  The negative cumulative gravitational disc
torque is achieved somewhat outside the orbit crossing radius. A small
amount of disc mass, less than 10\%, that is located outside that
radius provides the net negative torque on the disc. In that region of
space, $r \ga 0.41 r_{\rm H}$, the spiral arms are very prominent in
Figure~\ref{final} and gas response is highly nonlinear.  A similar
spiral structure was found in the case of a warm disc with $H/r=0.1$
within a binary star system by \cite{savonije94} that we have argued
should be similar to the circumplanetary disc case (see
Section~\ref{eqs}).

The SPH code is less accurate in these outer low density regions
because the interparticle pressure force calculation is more
approximate. Also, we have simplified the inflow on to the disc to be
occurring within an annulus in the disc.  But, we expect that the
general properties of gravitational torque to generally hold.  That
is, the gravitational torque involves a relatively small amount of gas
in the outer parts of disc.

\section{Discussion and Conclusions}
\label{concs}

\begin{figure*} 
\centering
\includegraphics[width=5.0cm]{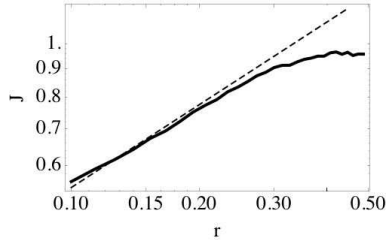}
\caption{Angular momentum per unit disc mass, $J$, in units of $ \Omega_{\rm p} r_{\rm H}^2$
  plotted against distance from the planet in units of $r_{\rm
    H}$ on a log-log scale. The solid line is  obtained from the
  SPH simulations described in Section 7 at a time of 6 orbits, while
  the dashed line is the Keplerian case. The solid curve is based on
  binning  in 30 radial bins, in order to smooth
  fluctuations.  For $r \simeq 0.3$,  inside of where
  orbits cross ($r \simeq 0.4$),  $J$ in the disc  departs from the Keplerian case
  due to the effects of gas pressure as the disc tapers. The outer
  disc velocity is somewhat subKeplerian because the pressure forces
  act radially outward. }
\label{vphi}
\end{figure*}

We have analysed the dynamics of a circumplanetary disc as an
accretion disc subject to the tidal forces of the central star. We
applied several techniques: ballistic particle orbits, one-dimensional
simulations and analytic models, and two-dimensional SPH simulations.
We have shown that the gas dynamical equations for a circumplanetary
disc can be rescaled to a form that is similar to those previously
analysed for a disc in a binary star system with mass ratio of unity,
but with a modified (Hill) potential.  The size of a circumplanetary
disc is determined by the requirement that tidal torques from the
central star remove angular momentum at a rate required by steady
inflow on to the planet. In the crudest approximation that is accurate
for a cold disc, disc streamlines consist of a set of stable, nested
periodic ballistic particle orbits (see Fig.~\ref{orbitcross}). For a
cold disk, the disc radius is determined to be the location of orbit
intersections or instability that occurs at $0.41 r_{\rm H}$, for Hill
radius $r_{\rm H}$.  This radius is close to the effective disc radius 
(inside of which $\sim$ 90\% of the mass is located)
found in the SPH simulations for a warm disc (see Figs.~\ref{final} and \ref{phase}).
In principle, the observational determination of the size of this disc then provides a constraint on the 
planet mass.

In some previous disc models, the angular momentum  of the inflowing
gas was thought to control the disc size. According to these models,
the compactness
of the giant planet satellite systems, within  a region of radius
much less than $r_{\rm H}$,  is a consequence of a low angular momentum inflow that penetrates
well inside the Hill sphere  and deposits presatellite material there \citep[e.g.][]{lissauer95, canup02, mosqueira03, ward10}.
We suggest  a different physical
model for the processes that control the circumplanetary disc
size and structure. We find that although the angular momentum per unit disc mass 
(specific angular momentum)  of the inflowing gas
plays a role in determining whether a circumplanetary disc can form,
it has a minor influence on the disc structure. A steady state accretion disc does not achieve as  compact a 
form as desired by  these previous models. These results imply
that the common scale of the disc seen in the simulations by \cite{ayliffe09} and
\cite{dangelo02} is not due to the specific angular momentum of the
inflow, but instead to the tidal truncation effects.  The angular
momentum within the disc is redistributed by viscous torques.  The
disc structure inside the truncation radius is mainly determined by
the properties of the disc turbulent viscosity, rather than the specific angular
momentum of the inflowing gas (see uppermost curves in Fig.~\ref{evolution}).
The average specific angular momentum in the disc is nearly independent
of the specific angular momentum of the accreting gas (see  Fig.~\ref{jc}).  
 Tidal effects have a large-scale influence on the disc density distribution, as discussed
in Section~\ref{analytic}.

We have concentrated on the case that the planet opens a gap in the disc, as is expected
for a Jupiter mass planet.
Some previous studies suggested that a small disc of size $\sim r_{\rm H}/48$
could result prior to and during early stages of gap opening \citep[e.g.][]{estrada08}.  Our results
suggest that a turbulent accretion disc would be truncated at the orbit crossing radius,
even if the inflowing gas has low specific angular momentum. There are
are few qualifications in this statement.
The simulations of \cite{ayliffe09}  found that circumplanetary discs 
do not exist prior to gap opening (planet masses less than 100 Earth masses). 
If low angular momentum gas is accreted in early stages of gap opening,
then pressure forces from within the Hill sphere could prevent a disc from extending
out to the orbit crossing radius, if the planet's Bondi radius is sufficiently smaller than
its Hill radius. Under such conditions, we would expect the disc
to expand to the Bondi radius by viscous torques, since tidal torques would
not be adequate to truncate the disc inside the Bondi radius. 
The disc would then need to 
lose its angular momentum to the surrounding gas in the circumstellar disc. 
Whether such a configuration is possible is unclear. In addition, gap opening or planet envelope
contraction (below its Hill sphere) might
not begin when the Bondi radius is smaller than the Hill radius and so a disc may not
form under such conditions.

During the T Tauri accretion phase, circumplanetary disc aspect ratios
are expected to be large, $H/r \sim 0.3$, as a consequence of the weak
gravitational forces due to the planet and relatively high
temperatures (see equation (\ref{Hr})).  There are some implications
of these high disc aspect ratios.  One is that the disc edge tapering
is more gradual than in the thin disc case. The tapering must be on a
scale $\ga H$ in order to avoid instabilities that occur for sharper
edges \citep[e.g.][]{yang10}. The tapering can then have an influence
on a substantial portion of the circumplanetary disc.  \cite{ayliffe09} estimated the disc outer
radius based on where the angular momentum per unit disc mass
departs from Keplerian and begins to
decline in radius. They obtained a disc outer radius of $\simeq 0.35
r_{\rm H}$ (see the upper right panel of their Fig. 2).  This radius
is somewhat inside the orbit crossing radius of $\simeq 0.41 r_{\rm  H}$.
The effect of
the disc tapering on the distribution of disc angular momentum in our SPH simulations
can be
seen in Fig.~\ref{vphi}. It leads to subKeplerian velocities inside
the orbit crossing radius, similar to the results of \cite{ayliffe09}.
The results of these two studies can  be further
reconciled by considering the differences in the models. In their
simulations, the disc is hotter than in our simulations. Consequently
the pressure forces and their effects on departures from Keplerian rotation in the outer parts
of the disc should be stronger than in Fig.~\ref{vphi}, as is consistent with their results. Also the work of
\cite{ayliffe09} includes the inflowing gas from larger radii that is
omitted in our simulations. This inflowing subKeplerian  gas and its interaction
with the disc outer edge as it becomes entrained could further modify
the results plotted in Fig.~\ref{vphi}.  It is also possible that the
disc radius is affected by the location of wave damping, as discussed
below.  Such effects should be explored further.

Another implication of the high values of $H/r$ is that a small fraction
of the disc mass can reside at larger disc radii than the orbit-crossing radius
and so be subject to the effects of resonances that lie outside
this radius, as
discussed in Section~\ref{res} (see Fig.~\ref{phase}). In addition,
off-resonant
tidal forcing of the outer parts of the disc that lie inside the orbit crossing radius can play an important
role. The resonance width depends on
$(H/r)^{2/3}$ and is not small. The substantial width can allow the resonant region to
overlap with the denser parts of disc. As a result of these two effects, two-armed spiral waves can be
launched in such discs, even though exact resonances do not lie within the main body of the discs. We have
found evidence of such waves in our SPH simulations (see
Fig.~\ref{final}).  For these waves to play a role in extracting
angular momentum from the disc, they must damp in order to introduce
irreversibility.  Otherwise, a standing wave is produced that results
in little or no torque on the disc. Some damping can be produced by
the disc turbulent viscosity in the fractional amount that is roughly
the ratio of the viscous wave damping rate to the wave group
propagation rate, about $\alpha (r/H)$. This ratio can be quite small
for $\alpha \ll H/r$ and the torque is reduced to a similarly small
fraction of its potentially maximum value.  For a thin disc, this
damping could occur as launched waves propagate, steepen and shock,
since their wavelengths are short compared to the disc radius.  But
for warm circumplanetary discs, this process is less important because
the wavelengths are not short compared to the disc radius.  In the SPH
simulations of Section 7, the damping seems to occur from the strongly
nonlinear forcing in the outer disc, similar to the case of mildly
warm discs ($H/r=0.1)$ previously investigated for binary star systems
\citep{savonije94}. The SPH simulations
suggest that the negative torque is produced in the outer parts of the disc, somewhat inside
and beyond the radius where nested particle orbits cross or become
unstable. This torque involves a relatively small amount of
disc mass (see Fig.~\ref{phase}).  However, the discs in our
SPH simulations were two-dimensional and not as warm as could occur, where
$H/r \sim 0.3$. The wave damping issue should be explored in future work.

The accretion disc model does not provide an obvious explanation for
the locations of the regular satellites of Jupiter and Saturn that
occur within $0.06 r_{\rm H}$ of the planet. As discussed above, the
disc structure is insensitive to the angular momentum of the inflowing
gas and satellites lie well inside the tidal truncation radius of the
disc. This conclusion is consistent with the results of recent
three-dimensional simulations of \cite{estrada08} and
\cite{ayliffe09}.  We have considered circumplanetary discs with a
smoothly varying turbulent viscosity with radius.  But such discs
could also harbor dead zones, as discussed in
Section~\ref{properties}.  Such situations could result in rapid disc
density variations at their boundaries. The existence of satellites
requires survival against the effects of migration.  Such density
variations could affect satellite migration and possibly trapping
satellites, since migration rates depend on density gradients
\citep[e.g.,][]{matsumura07}. But, if the inner dead zone boundary is
due to thermal ionisation at a temperature of $~10^3\,\rm K$, then the
disc temperatures there would be too hot for the survival of an icy
satellite. Of course, other sources of ionisation could change the
locations of dead zone edges.  In addition, the nonlinear feedback
from waves in a low viscosity disc could slow migration
\citep{ward97,rafikov03,li09}.  However, this slowing is less
effective for the warmer circumplanetary discs. If the satellite
formation occurs after partial disc depletion, as suggested by Canup
\& Ward (2002), then the ionisation through the vertical extent of the
disc becomes easier and the dead zones are less likely to occur at
that stage.

Circumplanetary discs are not very bright as we see in
equation~(\ref{lum}). Time dependent accretion and outbursts could
occur in dead zones, as has been suggested to explain FU Orionis
outbursts in young stellar systems \citep*{armitage01}.  During a
circumplanetary disc outburst, the ratio of the circumplanetary to
circumstellar luminosities could be much higher.

\section*{Acknowledgments}
We thank Matthew Bate,  Gennaro D'Angelo, and Jim Pringle  for helpful discussions.  RGM thanks the Space
Telescope Science Institute for a Giacconi Fellowship. SHL
acknowledges support from NASA grant NNX07AI72G.

\label{lastpage}
\end{document}